\documentclass[aps,preprint,showpacs,nofootinbib,tightenlines]{revtex4}

\usepackage{graphicx}  %
\usepackage{amsmath}
\usepackage{bm}  %
\usepackage{empheq}
\usepackage{ulem}

\newcommand{\bea}{\begin{eqnarray}}
\newcommand{\eea}{\end{eqnarray}}
\newcommand{\beq}{\begin{equation}}
\newcommand{\eeq}{\end{equation}}
\newcommand{\bse}{\begin{subequations}}
\newcommand{\ese}{\end{subequations}}

\def\mqo2{{\!\!\!}}

\begin{document}

\title{
Galilean-Invariant XEFT\\
at Next-to-Leading Order}

\author{Eric Braaten}
\email{braaten.1@osu.edu}
\affiliation{Department of Physics,
         The Ohio State University, Columbus, OH\ 43210, USA}

\author{Li-Ping He}
\email{he.1011@buckeyemail.osu.edu}
\affiliation{Department of Physics,
         The Ohio State University, Columbus, OH\ 43210, USA}

\author{Jun Jiang}
\email{jiangjun87@sdu.edu.cn}
\affiliation{School of Physics, Shandong University, Jinan, Shandong 250100, China}

\date{\today}

\begin{abstract}
XEFT is a low-energy effective field theory for charm mesons and pions
that provides a systematically improvable description of the $X(3872)$ resonance.
To simplify calculations  beyond leading order, we introduce  a new formulation of XEFT
with a dynamical field for a pair of charm mesons in the resonant channel.
We simplify the renormalization of XEFT by introducing a new renormalization scheme 
that involves the subtraction of amplitudes at the complex $D^{*0} \bar D^0$ threshold.
The new formulation and the new renormalization scheme are illustrated by calculating 
the complex pole energy of $X$ and the $D^{*0} \bar D^0$ scattering amplitude
to next-to-leading order using Galilean-invariant XEFT.
\end{abstract}

\smallskip
\pacs{14.40.Rt,14.40.Lb}
\keywords{$X(3872)$, effective field theory, Galilean invariance, charm mesons, pions}
\maketitle

\section{Introduction}
\label{sec:intro}

The $X(3872)$ was the first of the dozens of exotic hadrons whose fundamental constituents include
a heavy quark and its antiquark that have been discovered since the beginning of the century 
\cite{Ali:2017jda,Olsen:2017bmm,Karliner:2017qhf,Brambilla:2019esw}. 
Determining the nature of $X$ remains central to the problem of understanding the exotic heavy hadrons.
The discovery of $X$ by the Belle collaboration in 2003 was through the decay mode $J/\psi\, \pi^+\pi^-$,
which reveals that its constituents include a charm quark and its antiquark ($c \bar c$) \cite{Choi:2003ue}.
The $J^{PC}$ quantum numbers of $X$ were determined to be $1^{++}$ by the LHCb collaboration in 2013 \cite{Aaij:2013zoa}.
The possibilities for the particle structure of $X$ that are compatible with this information include
\begin{itemize}
\item 
the $\chi_{c1}(2P)$ charmonium state, whose quark constituents are $c \bar c$,
\item
a compact isospin-1 tetraquark meson, whose diquark constituents are
 $(c u) (\bar c \bar u) - (c d) (\bar c \bar d)$,
 \item
an isospin-0 charm-meson molecule, whose hadron constituents are
$(D^{*0} \bar D^0 + D^0 \bar D^{*0}) +(D^{*+}  D^- + D^+ D^{*-}) $,
which correspond to quark constituents $(c \bar u) (\bar c u) + (c \bar d) (\bar c d)$.
\end{itemize}

The Belle collaboration discovered $X$ at a mass that was surprisingly close to the scattering threshold for $D^{*0} \bar D^0$.
The measured energy relative to the threshold was $(+0.9 \pm 1.3)$~MeV \cite{Choi:2003ue}.
They put an upper bound on its width of about 2.3~MeV.
Over the subsequent years, the measurements  of the masses of $X$, $D^{*0}$, and $\bar D^0$
have all been improved significantly.
The LHCb collaboration has recently made the most precise measurements of the mass of $X$ to date,
and they made the first measurements of its width \cite{Aaij:2020qga,Aaij:2020xjx}.
With the line shape of the $X$ in the $J/\psi\, \pi^+\pi^-$ decay channel modeled by that of a Breit-Wigner resonance,
their results for the energy relative to the $D^{*0} \bar D^0$  threshold and for the width are
\bse
\bea
E_\mathrm{BW} &=& (-0.07 \pm 0.12)~\mathrm{MeV},
\label{EBW}
\\
\Gamma_\mathrm{BW} &=& (1.19 \pm 0.19)~\mathrm{MeV}.
\label{GammaBW}
\eea
\ese
An alternative prescription for the energy and the width of a resonance are the real and imaginary parts 
of the pole energy $E_X - i \Gamma_X/2$.
With the line shape modeled by that of a Flatt\'{e} amplitude that takes into account the width of the $D^{*0}$,
the LHCb collaboration obtained $(+0.025 -0.140\,i )$~MeV for the pole energy  
relative to the $D^{*0} \bar D^0$  threshold \cite{Aaij:2020qga}.
The energy $E_X$ is consistent with the measurement in Eq.~\eqref{EBW},
while the width $\Gamma_X$ is about 4 times smaller than the measurement in Eq.~\eqref{GammaBW}.

The extremely small energy of $X$ relative to the  $D^{*0} \bar D^0$  threshold in Eq.~\eqref{EBW}
has dramatic implications for the structure of $X$.
The quantum numbers $1^{++}$ of $X$ imply that it has an S-wave coupling to $D^{*0} \bar D^0$ and $D^0 \bar D^{*0}$.
Since these mesons are electrically neutral, they interact with each other through short-range interactions 
with a range of order $1/m_\pi$, where $m_\pi$ is the pion mass.
However the tiny energy of $X$ relative to the  $D^{*0} \bar D^0$  threshold 
implies that the  interaction between the charm mesons is resonant.
Thus $X$ is an S-wave resonance near the threshold for a pair of particles with short-range interactions.
General principles of quantum mechanics guarantee that $X$ must have universal properties 
determined by its binding energy \cite{Braaten:2003he}.
They guarantee that the dominant component of $X$ must be a charm-meson molecule with the particle structure
\beq
X =  \frac{1}{\sqrt{2}} \big( D^{*0} \bar D^0 + D^0 \bar D^{*0} \big).
\label{X-DD}
\eeq
If $E_X<0$, $X$ is a bound state whose  spatial structure   is described by a universal wavefunction $\exp(-r/a)/r$,
where $a=1/\sqrt{2\mu |E_X|}$ is the S-wave scattering length of $D^{*0} \bar D^0$ in the $C=+$ channel
and $\mu$ is the reduced mass of $D^{*0} \bar D^0$.
The mean separation of the constituents is $\langle r\rangle = a/2$.
If the energy $E_X$ is identified with the Breit-Wigner energy in Eq.~\eqref{EBW},
the lower bound on the energy at the 90\% confidence level  is $E_X > -0.22$~MeV.
Thus the mean separation is larger than 4.8~fm.

The universality of S-wave near-threshold resonances is a truly remarkable aspect 
of quantum mechanics \cite{Braaten:2004rn}.
If a model for the $X$ with any particle structure, such as one of those itemized above,
 is extended to allow couplings to $D^{*0} \bar D^0$  and $D^0 \bar D^{*0}$ scattering states,
 it will be dramatically transformed by the resonant couplings to the scattering states.
 If the model has an adjustable parameter that can be used to tune the resonance energy to the
$D^{*0} \bar D^0$ threshold, the resonance will in the limit develop the particle structure in Eq.~\eqref{X-DD}
with the universal wavefunction $\exp(-r/a)/r$.
This remarkable phenomenon is widely recognized in the case of a charm-meson molecule.
That it occurs also in the case of a charmonium state or a compact tetraquark is not
as widely recognized.

The universality of S-wave near-threshold resonances provides a basis for a systematically improvable
treatment of the $X$ resonance using effective field theory.
An appropriate effective field theory was invented by Fleming {\it et al.}\ and named XEFT \cite{Fleming:2007rp}.
XEFT is a nonrelativistic effective field theory for charm mesons and pions.
In the simplest version of XEFT, the only fields are those for the neutral charm mesons 
$D^{*0}$, $\bar D^{*0}$, $D^0$, and $\bar D^0$ and the neutral pion $\pi^0$.
The only components of $X$ that are treated explicitly are those in Eq.~\eqref{X-DD} and $D^0 \bar D^0 \pi^0$.
The effects of all other particles must be taken into account in the parameters of XEFT.
In particular, different models for $X$ with resonance energy far enough away from the $D^{*0} \bar D^0$ threshold,
such as those itemized above, correspond to different choices for the parameters of XEFT.
The simplest version of XEFT is sufficient if the total energy
of $D^{*0} \bar D^0$, $D^0 \bar D^{*0}$, or  $D^0 \bar D^0 \pi^0$ is close enough to the $D^{*0} \bar D^0$ threshold.
The region of validity of XEFT can be extended by adding  fields 
for the charged charm mesons $D^{*+}$, $D^{*-}$, $D^+$, and $D^-$ and the charged pions $\pi^+$ and $\pi^-$.
In this case, the additional components of $X$ that are treated explicitly are 
 $D^{*+} D^-$, $D^+ D^{*-}$, $D^+ D^- \pi^0$, $D^+ \bar D^0 \pi^-$, and $D^0 D^- \pi^+$.

An effective field theory can be simplified by taking advantage of exact and approximate symmetries.
A remarkable aspect of the sector of QCD consisting of $D^* \bar D$, $D \bar D^*$, and $D \bar D \pi$ 
that is described by XEFT is that  the sum of the masses is very nearly conserved.
Galilean invariance is a possible  symmetry of a nonrelativistic field theory that  
requires exact conservation of the kinetic mass.
In Ref.~\cite{Braaten:2015tga}, a Galilean formulation of XEFT was developed.
Galilean invariance provides strong constraints on the ultraviolet (UV) divergences from loop amplitudes.
It therefore greatly simplifies the renormalization of XEFT.
In this paper, we introduce a new formulation of Galilean-invariant XEFT with a dynamical field  
for a pair of charm mesons in the resonant channel.
This new formulation further simplifies calculations beyond leading order.

The accuracy of an effective field theory can be greatly improved by using an appropriate renormalization scheme.
Analytic results can also be greatly simplified by the choice of an appropriate renormalization scheme.
In the pioneering paper on XEFT, the momentum distribution for the decay of $X$ into $D^0 \bar D^0 \pi^0$ 
was calculated at next-to-leading order (NLO) using dimensional regularization 
with power divergence subtraction \cite{Fleming:2007rp}.
The only UV divergences are linear divergences, and they were removed by absorbing them into interaction parameters.
In Ref.~\cite{Braaten:2015tga}, the elastic scattering amplitude for $D^{*0} \bar D^0$ 
was calculated   at NLO using dimensional regularization in Galilean-invariant  XEFT. 
There are both linear and logarithmic UV divergences,
and they were removed by subtractions at the complex pole energy of $X$.
In this paper, we introduce a simpler renormalization scheme for XEFT 
in which divergences are removed instead by subtractions at the complex threshold energy of $D^{*0} \bar D^0$.
This new renormalization scheme greatly simplifies analytic results at NLO.

In Section~\ref{sec:newXEFT},  we introduce Galilean invariance and
we describe various  formulations of XEFT. 
In Section~\ref{sec:Lagrangian}, we present the Lagrangian for  the new formulation of 
Galilean-invariant XEFT  with a dynamical pair field.
In Section~\ref{sec:Frules}, we present the Feynman rules for  this new formulation of XEFT.
In Section~\ref{sec:PairPropNLO}, we calculate the pair propagator at NLO 
and we obtain the complex pole energy of $X$ at NLO.
In Section~\ref{sec:D*Dscat},  we calculate the $D^{*0} \bar D^0$ elastic scattering amplitude at NLO
and we analyze the breakdown of the effective range expansion from pion exchange.
We summarize our results and suggest other useful applications in Section~\ref{sec:Outlook}.
In Appendix~\ref{sec:LoopInt}, we present results for loop integrals that arise in  calculations at NLO.
In Appendix~\ref{sec:NLOAmp}, we present the results for individual Feynman diagrams
for the $D^{*0} \bar D^0$  transition amplitude at NLO.

\section{Galilean-invariant XEFT}
\label{sec:newXEFT}

In this section, we introduce Galilean invariance,
describe various formulations of XEFT, and give the numerical values of some of its parameters.

\subsection{Galilean invariance}
\label{sec:galilean}

{\it Galilean invariance} is a possible space-time symmetry 
of a nonrelativistic field theory  \cite{Hagen:1972pd,Rosen:1972sh}. 
In a nonrelativistic theory, the energy-momentum relation for an on-shell particle
with rest energy  $\varepsilon$ and kinetic mass $m$ is
\beq	
E(p) = \varepsilon + p^2/(2m).
\label{E-p:NR}
\eeq
Galilean symmetry requires invariance under a Galilean boost with an arbitrary velocity vector $\bm{v}$.
The effects of the Galilean boost on the energy and momentum are
\begin{subequations}
\bea
E &\longrightarrow&E+ \bm{v} \cdot \bm{p} + \frac12 m v^2,
\label{boost:E}
\\
\bm{p} &\longrightarrow& \bm{p} + m  \bm{v}.
\label{boost:p}
\eea
\label{boost}%
\end{subequations}
The  {\it invariant energy} $E-p^2/2m$ is invariant under Galilean boosts.

A unique feature of the sector of QCD consisting of spin-0 charm mesons $D$ and $\bar D$, 
spin-1 charm mesons $D^*$ and $\bar D^*$, and pions $\pi$
is that mass is very nearly conserved by the transitions $D^* \leftrightarrow D \pi$.
In the decay $D^{*0} \to D^0 \pi^0$, the sum of the masses of the $D^0$ and $\pi^0$ 
is  lower than the mass of the $D^{*0}$ by 7.0~MeV, which is only 0.35\%.
Since the isospin splittings  between the charm mesons are at most 4.8~MeV 
and the isospin splitting between  $\pi^+$ and $\pi^0$ is 4.6~MeV,
all the transitions $D^* \leftrightarrow D \pi$ come very close to satisfying mass conservation.
There are no other hadrons with such narrow widths 
that have transitions that come so close to satisfying mass conservation.

Galilean invariance requires the exact conservation of the kinetic mass \cite{Hagen:1972pd}.
In a Galilean-invariant description of charm mesons and pions, 
the spin-0 charm mesons must all have the same kinetic mass $M$
and the pions must all have the same kinetic mass $m$.
Conservation of kinetic mass then requires the kinetic mass of all the spin-1 charm mesons to be $M+m$.

In a nonrelativistic effective field theory for charm mesons, 
one can impose a phase symmetry that guarantees the separate conservation 
of the number $N_c$ of charm quarks 
and the number $N_{\bar c}$ of charm antiquarks.
These quark numbers can be expressed in terms of meson numbers.
In a theory with only neutral charm mesons and $\pi^0$, the quark numbers are
\begin{subequations}
\bea
N_c &=& N_{D^{*0}} + N_{D^0},
\label{Nc}
\\
N_{\bar c} &=& N_{\bar D^{*0}} + N_{\bar D^0}.
\label{Ncbar}
\eea
\label{Nccbar}%
\end{subequations}
If the theory also includes  charged charm mesons and charged pions,
the charm quark number $N_c$  also includes the numbers of $D^{*+}$ and $D^+$.
In a Galilean-invariant effective field theory for charm mesons and pions,
the exact conservation of kinetic mass in the transitions $D^* \leftrightarrow D \pi$ provides motivation for introducing 
an additional phase symmetry that guarantees the conservation of {\it pion number}.
In a theory with only neutral charm mesons and $\pi^0$, the pion number is
\beq	
N_\pi = N_{\pi^0} + N_{D^{*0}} + N_{\bar D^{*0}}.
\label{Npi}
\eeq
If  the theory also includes charged charm mesons and charged pions,
the pion number $N_\pi$ also includes the numbers of $\pi^+$, $\pi^-$, $D^{*+}$, and $D^{*-}$.

\subsection{XEFT}
\label{sec:XEFT}

XEFT is a nonrelativistic effective field theory for charm mesons and pions invented by Fleming {\it et al.}\ \cite{Fleming:2007rp}.
It provides a systematically improvable description of the sector of QCD consisting of 
$D^*\bar D$, $D \bar D^*$, and $D \bar D \pi$ with total energy  near the $D^*\bar D$ threshold.
It therefore can be used to calculate some properties of the $X$ resonance systematically.
In XEFT with only neutral charm mesons and $\pi^0$,
the only  fields are complex scalar fields for $D^0$, $\bar D^0$, and  $\pi^0$
and complex vector fields for $D^{*0}$ and $\bar D^{*0}$.
For the  sector of QCD consisting of $D^{*0} \bar D^0$, $D^0 \bar D^{*0}$, $D^0 \bar D^0 \pi^0$, and $X$,
the region of validity of XEFT with only neutral particles extends at most to the $D^{*+} D^-$  threshold,
which is 8.2~MeV above the $D^{*0}\bar D^0$ threshold.
The region of validity can be extended to higher energies by introducing additional fields 
for the charged charm mesons and the charged pions.

At leading order (LO) in the power counting of XEFT,
the only adjustable parameter is the LO binding momentum $\gamma$ of $X$,
which was assumed to be a real parameter in Ref.~\cite{Fleming:2007rp}.
At next-to-leading order (NLO), there are additional adjustable interaction parameters.
In Ref.~\cite{Fleming:2007rp}, XEFT was used to calculate the differential decay rate of $X$ into $D^0 \bar D^0 \pi^0$
at  NLO.   It depends on two additional adjustable real parameters:
a length $r_0$ associated with the  effective range in the resonant S-wave even-charge-conjugation ($C=+$) channel
and a parameter for the coupling of $D^{*0}\bar D^0$ to $D^0 \bar D^0 \pi^0$.
Fleming {\it et al.}\ calculated the partial decay rate of $X$  into $D^0 \bar D^0 \pi^0$ numerically 
as a function of the LO binding energy $\gamma^2/(2 \mu)$.  
Their estimate of the width of the error band from the two NLO interaction parameters
decreased from about 25\% to about 10\% as $\gamma^2/(2 \mu)$ decreased from 0.1~MeV to 0.01~MeV.
The relatively  wide  error bands even for extremely tiny values of $\gamma$
raises the question 
of whether the power-counting expansion for XEFT converges fast enough for it to be quantitatively useful.
The original formulation of XEFT in Ref.~\cite{Fleming:2007rp}
has also been applied to the $D^{*0}\bar D^0$ scattering length at NLO \cite{Jansen:2013cba}.
The NLO  calculation of the decay rate for $X$ into $D^0 \bar D^0 \pi^0$
was recently revisited by Dai {\it et al.}\ \cite{Dai:2019hrf}.
They pointed out that the power-counting rules of chiral effective field theories for heavy mesons implied 
that there are two additional NLO interactions parameters:
the S-wave scattering lengths for $D^0 \bar D^0$ and for $D^0 \pi^0$.
Their estimate of the width of the error band from the four NLO interaction parameters
decreased from about 60\% to about 30\% as $\gamma^2/(2 \mu)$ decreased from 0.1~MeV to 0.01~MeV.
The wide error bands even for extremely tiny values of $\gamma$
further emphasizes the convergence problem of XEFT.

An alternative  formulation of XEFT was developed in Ref.~\cite{Braaten:2015tga}.
It differs from the original formulation in Ref.~\cite{Fleming:2007rp} in three important ways:
\begin{enumerate}
\item
Galilean invariance.
\item
Systematic treatment of the width of $D^{*0}$,
which requires the LO binding momentum $\gamma$ to be complex.
\item
Complex on-shell  (COS)  renormalization scheme,  in which the  UV divergences in 
$D^{*0}\bar D^0$ and $D^0 \bar D^{*0}$  amplitudes 
are removed by subtractions at the complex pole energy $E_X - i \Gamma_X/2$.
\end{enumerate}
Taking into account the $D^{*0}$ width is essential if XEFT is to give an accurate  description of the $X$ resonance.
In the COS scheme, the adjustable real parameters at LO are $E_X$ and $\Gamma_X$.
In Ref.~\cite{Braaten:2015tga},
Galilean-invariant XEFT was used to calculate the $D^{*0}\bar D^0$ scattering length at NLO. 
The result depends on a single additional adjustable parameter 
that can be identified with the parameter $r_0$ in Ref.~\cite{Fleming:2007rp}.
One drawback of the COS scheme is that the analytic expression for 
the $D^{*0}\bar D^0$ scattering amplitude is rather complicated.
Galilean invariance simplifies the renormalization of XEFT. 
It implies, for example, that the S-wave scattering lengths for $D^0 \bar D^0$ and for $D^0 \pi^0$, 
which were argued to be NLO parameters of XEFT in Ref.~\cite{Dai:2019hrf}, 
are not required by renormalization.

In this paper, we present an alternative formulation of Galilean-invariant XEFT. 
It differs from the formulation of Galilean-invariant XEFT in Ref.~\cite{Braaten:2015tga} in three important ways:
\begin{enumerate}
\item
An additional complex vector field, which we call the {\it pair field}, 
that annihilates a pair of charm mesons in the resonant channel.
\item
New Feynman rules in which $D^{*0}\bar D^0$ and $D^0 \bar D^{*0}$ couple to the resonant channel 
only through the intermediate pair propagator.
\item
Complex threshold  (CT)  renormalization scheme,  in which the UV divergences in 
$D^{*0}\bar D^0$ and $D^0 \bar D^{*0}$ amplitudes 
are removed by subtractions at the complex threshold for $D^{*0}\bar D^0$ scattering states.
\end{enumerate}
The new Feynman rules simplify calculations beyond LO by  making some cancellations 
of UV divergences between diagrams automatic.
The CT scheme dramatically simplifies analytic expressions for amplitudes beyond LO.
It  reveals the existence of an additional adjustable interaction parameter at NLO 
that was not recognized in Ref.~\cite{Braaten:2015tga}.
It may also provide a solution to the problem of the large NLO corrections in XEFT 
that were encountered in Refs.~\cite{Fleming:2007rp,Dai:2019hrf}.

\subsection{Known parameters}
\label{sec:params}

We denote the masses of $D^{*0}$, $D^0$, and $\pi^0$ by $M_{*0}$, $M_0$, and $m_0$, respectively.
The difference between the $D^{*0}$ mass and the $D^0 \pi^0$ threshold is
\begin{equation}
\delta \equiv M_{*0} - (M_0 + m_0) = (7.04 \pm 0.03)~\mathrm{MeV}.
\label{delta-M*Mm}
\end{equation}
The decay width of the $D^{*0}$ can be predicted by assuming  the decays $D^* \to D \pi$ respect 
chiral symmetry, isospin symmetry, and Lorentz invariance.
The measured branching fraction for $D^{*0} \to D^0 \pi^0$ is $B_{*0\pi}= (64.7\pm0.9)\%$.
Using  the decay width of the $D^{*+}$ and the branching fractions for $D^{*0} \to D^0 \pi^0$
and $D^{*+} \to D^0 \pi^+$ as  inputs, the prediction for the total $D^{*0}$ width is
\begin{equation}
\Gamma_{*0} \equiv \Gamma[D^{*0}] = (55.4 \pm 1.5)~\mathrm{keV}.
\label{eq:Gamma*0}
\end{equation}

In Galilean-invariant XEFT, the spin-0 charm mesons all have the same kinetic mass $M$
and the pions all have the same kinetic mass $m$.
We choose the kinetic masses $M$ and $m$ of $D^0$ and $\pi^0$
 to be equal to their physical masses $M_0$ and $m_0$, respectively.
The rest energies of  $D^0$ and $\pi^0$ are therefore both zero.
Galilean invariance requires the kinetic mass of $D^{*0}$ to be $M+m$.
The  $D^{*0}$ has the  complex rest energy
\beq	
E_* = \delta -i \Gamma_{*0}/2,
\label{E*}
\eeq
where $\delta$ is the energy in Eq.~\eqref{delta-M*Mm} 
and $\Gamma_{*0}$ is  the $D^{*0}$ width  in Eq.~\eqref{eq:Gamma*0}.

The reduced  mass $\mu_\pi$ for $D^0 \pi^0$  and the reduced kinetic mass $\mu$ for $D^{*0} \bar D^0$  are
\begin{subequations}
\bea
\mu_\pi &\equiv& \frac{mM}{M+m} = 125.87~{\rm MeV},
\label{mupi}
\\
\mu &\equiv& \frac{M(M+m)}{2M+m} = 965.0~{\rm MeV}.
\label{mu*}
\eea
\label{mumu}%
\end{subequations}
The ratio of these reduced masses is a small parameter in Galilean-invariant XEFT:
\beq
r \equiv \mu_\pi/\mu= 0.1304.
\label{r-mu}
\eeq

The $D^{*0}$-to-$D^0\pi^0$ coupling constant  in XEFT is conventionally denoted by $g/(2\sqrt{m} f_\pi)$.
This coupling constant can be determined from the partial width of $D^{*0}$ into $D^0 \pi^0$:
\beq	
\Gamma_{*0,\pi} \equiv \Gamma[D^{*0} \to D^0 \pi^0] =
\left( \frac{g^2}{4m f_\pi^2} \right)
\frac{\mu_\pi}{3 \pi}  ( 2 \mu_\pi \delta)^{3/2},
\label{GammaD*Dpi0}
\eeq
where $\delta$ is the energy in Eq.~\eqref{delta-M*Mm}.
It is convenient to define a coupling constant $g_\pi^2$ with dimensions 1/(momentum):
\beq	
g_\pi^2 \equiv (g^2/4 m f_\pi^2) \mu^2 =
 \big[ (29.7 \pm 0.7)~\mathrm{MeV} \big]^{-1}.
\label{gpi2}
\eeq
%

\section{Lagrangian}
\label{sec:Lagrangian}

In this section, we write down the Lagrangian for Galilean-invariant XEFT
for the neutral charm mesons and $\pi^0$.
We introduce a new form for the Lagrangian with a dynamical pair field
for  a pair of charm mesons in the resonant channel.
We include all terms required to calculate to NLO in the XEFT power counting.

\subsection{LO Lagrangian}
\label{sec:LagrangianLO}

The fields for the $D^0$ and $\bar D^0$ are complex scalar fields $D$ and $\bar D$. 
The field for the $\pi^0$ is a complex scalar field $\pi$. 
The fields for the $D^{*0}$ and $\bar D^{*0}$ are complex vector fields $\bm{D}$ and $\bar{\bm{D}}$. 
The kinetic terms in the Lagrangian for $D^0$ and $\pi^0$ are
\begin{subequations}
\bea
{\cal L}_D &=& D^\dagger \left[ i \partial_t + \nabla^2/(2M) \right] D,
\label{LD}
\\
{\cal L}_\pi &=& \pi^\dagger \left[ i \partial_t + \nabla^2/(2m) \right] \pi.
\label{Lpi}
\eea
\label{Lkinetic}%
\end{subequations}
The kinetic term for $\bar D^0$ is obtained  from Eq.~\eqref{LD} by replacing $D$ by $\bar D$. 
The kinetic term in the Lagrangian for $D^{*0}$ is
\beq
{\cal L}_{D^*} = \bm{D}^\dagger \cdot \left[ i \partial_t + \nabla^2/(2(M+m))  - E_*\right] \bm{D},
\label{LD*}
\eeq
where $E_*$ is the complex rest energy of $D^{*0}$ in Eq.~\eqref{E*}.
The kinetic term for $\bar D^{*0}$ is obtained by replacing  $\bm{D}$ by $\bar{\bm{D}}$.

There is a  resonance in the S-wave channel  for the superposition 
of $D^{*0} \bar D^0$ and $D^0 \bar D^{*0}$ with  even charge conjugation ($C=+$) in Eq.~\eqref{X-DD}.
The interaction term in the Lagrangian for XEFT at LO is a contact interaction in the resonant channel:
\beq
{\cal L}_{\mathrm{LO,int}} = - \frac{C_0}{2} \,
\left(  \bar  D \bm{D} +    D \bar{\bm{D}}\right)^\dagger \cdot \left(  \bar  D \bm{D} +    D \bar{\bm{D}}\right).
\label{L-LOint}
\eeq
It is convenient to introduce a complex vector field $\bm{\phi}$ that we call the 
{\it pair field}  that annihilates a pair of charm mesons in the resonant channel:
\beq
\bm{\phi} = \frac{C_0}{\sqrt{2}} \, \left(  \bar  D \bm{D} +    D \bar{\bm{D}}\right) .
\label{Phi}
\eeq
The normalization factor  has been chosen for later convenience.
Using the pair field $\bm{\phi}$, we can write down an alternative interaction term in the Lagrangian for XEFT at LO:
\beq
{\cal L}_{\mathrm{LO,int}} = \frac{1}{C_0} \, \bm{\phi}^\dagger \cdot  \bm{\phi}
- \frac{1}{\sqrt{2}} \left[\left(  \bar  D \bm{D} +    D \bar{\bm{D}}\right)^\dagger \cdot \bm{\phi} 
+ \bm{\phi}^\dagger \cdot \left(  \bar  D \bm{D} +    D \bar{\bm{D}}\right) \right].
\label{L-LOint:Phi}
\eeq
The field equation  for $\bm{\phi}$ implies Eq.~\eqref{Phi}. 
The field equation can be used to eliminate $\bm{\phi}$ from Eq.~\eqref{L-LOint:Phi}, which reduces it to Eq.~\eqref{L-LOint}.
The two interaction Lagrangians are therefore equivalent.

\subsection{NLO interaction terms}
\label{sec:LagrangianNLO}

In the  original paper on XEFT in Ref.~\cite{Fleming:2007rp}, 
all the interaction terms in the Lagrangian  needed for calculations 
to next-to-leading order (NLO) in the power-counting of XEFT were written down explicitly.
They included pion interaction terms that allow transitions between $D^{*0}$ and $D^0\pi^0$ 
and between $\bar D^{*0}$ and  $\bar D^0 \pi^0$.
In Galilean-invariant XEFT, the pion interaction terms  are \cite{Braaten:2015tga} 
\beq	
{\cal L}_{D^* \leftrightarrow D \pi} =
\frac{g}{2\sqrt{m} f_\pi} 
\left[ \bm{D}^\dagger \cdot 
\big(D \overleftrightarrow{\bm{\nabla}} \pi \big)
+ \big(D  \overleftrightarrow{\bm{\nabla}} \pi \big)^\dagger \cdot \bm{D}
+  \bar{\bm{D}}^\dagger \cdot 
\big(\bar{D} \overleftrightarrow{\bm{\nabla}} \pi \big)
+ \big(\bar{D}  \overleftrightarrow{\bm{\nabla}} \pi \big)^\dagger
\cdot \bar{\bm{D}} \right],
\label{Lintpi}
\eeq
where $\overleftrightarrow{\bm{\nabla}}=(M \overrightarrow{\bm{\nabla}} - m \overleftarrow{\bm{\nabla}})/(M+m)$
is a Galilean-invariant derivative.
The pion interaction term in  the Lagrangian for original XEFT in Ref.~\cite{Fleming:2007rp}
can be obtained by replacing the operator $\overleftrightarrow{\bm{\nabla}}$
in Eq.~\eqref{Lintpi}  by $\overrightarrow{\bm{\nabla}}$.

The NLO interaction terms in the Lagrangian for original XEFT in Ref.~\cite{Fleming:2007rp} include
$\nabla^2$ interaction terms that produce transitions between incoming $D^{*0} \bar D^0$ or $D^0 \bar D^{*0}$ 
and outgoing $D^{*0} \bar D^0$ or $D^0 \bar D^{*0}$.
There are equivalent interaction terms  involving the  pair field $\bm{\phi}$.
In Galilean-invariant XEFT, the $\nabla^2$ interaction terms involving $\bm{\phi}$  are
\bea
{\cal L}_{\nabla^2} &=& 
 \frac{C_2}{2\sqrt2\,  C_0}
\left[ \bm{\phi}^\dagger \cdot 
\Big(\bar D \overleftrightarrow{\bm{\nabla}}^2 \bm{D} 
+ D \overleftrightarrow{\bm{\nabla}}^2\bar{\bm{D}} \Big)
+ \Big(\bar D \overleftrightarrow{\bm{\nabla}}^2 \bm{D} 
+ D \overleftrightarrow{\bm{\nabla}}^2\bar{\bm{D}} \Big)^\dagger \cdot  \bm{\phi} \right] ,
\label{Lnabla2}
\eea
where $\overleftrightarrow{\bm{\nabla}}=(M \overrightarrow{\bm{\nabla}} - (M+m) \overleftarrow{\bm{\nabla}})/(2M+m)$
is a Galilean-invariant derivative.
The  $\nabla^2$ interaction  terms in Ref.~\cite{Braaten:2015tga} can be obtained 
by eliminating $\bm{\phi}$ using the field equation in Eq.~\eqref{Phi}.
The $\nabla^2$ interaction terms in the Lagrangian for original XEFT in Ref.~\cite{Fleming:2007rp}
can be obtained by replacing the operator $\overleftrightarrow{\bm{\nabla}}$ in Eq.~\eqref{Lnabla2} by  
$(\overrightarrow{\bm{\nabla}} - \overleftarrow{\bm{\nabla}})/2$
and then eliminating $\bm{\phi}$ using its field equation.

The NLO interaction terms  in the Lagrangian for original XEFT in Ref.~\cite{Fleming:2007rp} include
counterterms  that produce transitions between incoming $D^{*0} \bar D^0$ or $D^0 \bar D^{*0}$ 
and outgoing $D^{*0} \bar D^0$ or $D^0 \bar D^{*0}$.
There is an equivalent counterterm involving the  pair field only: 
\beq
{\cal L}_\mathrm{counter} =
- (1/C_0^2) \big[\delta C_0 \, \bm{\phi}^\dagger \cdot  \bm{\phi}
+D_0 \,  \bm{\phi}^\dagger \cdot ( i  D_t - E_*)  \bm{\phi} \big].
\label{Lct}
\eeq
where $i D_t= i \partial_t + \nabla^2/(2(2M+m))$ is a Galilean-invariant derivative. 
The constant subtracted from $i D_t$ is arbitrary. 
The choice $E_*$ corresponds to the complex on-shell renormalization scheme for the $D^{*0}$ propagator.
The  NLO counterterms  in the Lagrangian for original XEFT in Ref.~\cite{Fleming:2007rp} 
can be obtained by eliminating $\bm{\phi}$ from Eq.~\eqref{Lct} using the field equation in Eq.~\eqref{Phi}.

In the  original paper on XEFT in Ref.~\cite{Fleming:2007rp}, 
the Lagrangian included another NLO interaction term that allowed transitions 
between $D^{*0} \bar D^0$ or $D^0 \bar D^{*0}$ and $D^0 \bar D^0 \pi^0$.
There is an equivalent NLO interaction term  involving the pair field.
In Galilean-invariant XEFT, the $D \bar D \pi$ interaction terms involving $\bm{\phi}$ are
\beq
{\cal L}_{D \bar D \pi} =
\frac{B_1}{\sqrt{2m}\, C_0} 
\left[ \bm{\phi}^\dagger \cdot \big( D\bar{D} \overleftrightarrow{\bm{\nabla}} \pi \big) 
+ \big(D\bar{D} \overleftrightarrow {\bm{\nabla}}\pi \big) ^\dagger \cdot \bm{\phi} \right],
\label{L-DDpi}
\eeq
where $\overleftrightarrow{\bm{\nabla}}=(2M \overrightarrow{\nabla}-m \overleftarrow{\nabla})/(2M+m)$
is a Galilean-invariant derivative.
The $D \bar D \pi$ interaction term in  the Lagrangian for original XEFT in Ref.~\cite{Fleming:2007rp}
can be obtained by first replacing the operator $\overleftrightarrow{\bm{\nabla}}$
in Eq.~\eqref{L-DDpi}  by $\overrightarrow{\bm{\nabla}}$
and then eliminating $\bm{\phi}$ from Eq.~\eqref{L-DDpi} using the field equation in Eq.~\eqref{Phi}.

The NLO  interaction terms in Eqs.~\eqref{Lnabla2}, \eqref{Lct}, and \eqref{L-DDpi}
are all required by renormalization.
In the  original paper on XEFT in Ref.~\cite{Fleming:2007rp}, 
the momentum distributions from the decays of $X$ into $D^0 \bar D^0\pi^0$ were calculated to NLO. 
The ultraviolet divergences were removed by renormalizations of the 
coupling constants $C_2$ in Eq.~\eqref{Lnabla2} and $B_1$ in Eq.~\eqref{L-DDpi}.
In Ref.~\cite{Braaten:2015tga}, Galilean-invariant XEFT was used to calculate the 
$D^{*0} \bar D^0$ elastic scattering amplitude to NLO.
The ultraviolet divergences were removed by renormalizations of the 
coefficients $\delta C_0$ and $D_0$ in the counterterm in Eq.~\eqref{Lct}
and the coupling constant $C_2$ in Eq.~\eqref{Lnabla2}.

In the  original paper on XEFT in Ref.~\cite{Fleming:2007rp}, the authors developed power-counting rules.
A convenient way to implement the XEFT power counting is to assign orders in the coupling constant $g$ 
for pion emission and absorption to the coupling constants of all other interaction terms.
A complete calculation then requires calculating all diagrams to a given order in $g$.
The coupling constant $C_0$ is order $g^0$.
The coupling constant $C_2$ in Eq.~\eqref{Lnabla2} is order $g^2$.
The counterterm coefficients  $\delta C_0$ and  $D_0$ in Eq.~\eqref{Lct} are order $g^2$.
The coupling constant $B_1$ in Eq.~\eqref{L-DDpi} is order $g^3$.
These are the only coupling constants required by renormalization in XEFT at NLO.

In Ref.~\cite{Dai:2019hrf}, the authors argued that the Lagrangian for XEFT at NLO should also
include interactions terms that allow the S-wave scattering reactions $D^0 \bar D^0 \to D^0 \bar D^0$ and
$D^0 \pi^0  \to D^0 \pi^0 $.  These interaction terms are NLO 
according to the power-counting rules for chiral effective field theories of heavy mesons.
In Galilean-invariant XEFT, there are no diagrams for the reaction $D^0 \bar D^0 \to D^0 \bar D^0$, 
because the pion interaction terms in Eq.~\eqref{Lintpi}
do not allow an incoming $D^0$ or $\bar D^0$ to emit a $\pi^0$.
There are therefore no ultraviolet divergences that would require introducing an S-wave $D^0 \bar D^0$ interaction term.
In Galilean-invariant XEFT, the reaction $D^0 \pi^0  \to D^0 \pi^0 $ 
proceeds only in the P-wave channel through an intermediate $D^{*0}$. 
There are no loop diagrams, so there are no ultraviolet divergences that would require introducing an S-wave 
$D^0 \pi^0$ interaction term.
Thus there are no S-wave $D^0 \bar D^0$ and $D^0 \pi^0$ interaction
terms in Galilean-invariant XEFT.
These interactions terms  can of course be introduced as first-order perturbations to the Lagrangian of XEFT
 in order to estimate the effects of such hadronic interactions on observables.

\section{Feynman Rules}
\label{sec:Frules}

In this Section, we write down the Feynman rules for Galilean-invariant XEFT 
with a dynamical pair field, whose Lagrangian is given in Section~\ref{sec:Lagrangian}. 
A set of Feynman rules that are useful for calculations at NLO are given below as expressions enclosed in boxes.

\subsection{Particle propagators}
\label{sec:Props}

\begin{figure}[tb]
\centerline{\includegraphics*[width=5cm,angle=0,clip=true]{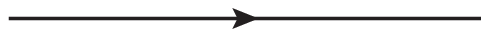}
\hspace{1cm}
\includegraphics*[width=5cm,angle=0,clip=true]{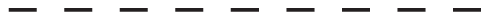}}
\vspace*{0.0cm}
\caption{
The  propagators for $D^0$ and $\pi^0$
are represented by a solid line with an arrow and a dashed line, respectively.  
The Feynman rules for these propagators are given in Eqs.~\eqref{propDpi}.
The propagator for $\bar D^0$ looks like that for $D^0$  with the arrow reversed.\\
}
\label{fig:propDpi}
\end{figure}

\begin{figure}[tb]
\centerline{\includegraphics*[width=5cm,angle=0,clip=true]{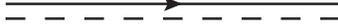}}
\vspace*{0.0cm}
\caption{
The  propagator for $D^{*0}$ is represented by a double line (solid with an arrow and dashed).
The Feynman rule for this propagator is given in Eq.~\eqref{propD*}.
The propagator for $\bar D^{*0}$ looks like that for $D^{*0}$ with the arrow reversed.\\
}
\label{fig:propD*}
\end{figure}

\begin{figure}[tb]
\centerline{\includegraphics*[width=5cm,angle=0,clip=true]{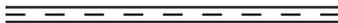}}
\vspace*{0.0cm}
\caption{
The  pair propagator is represented by a triple line (solid, dashed, and solid). 
The Feynman rule for this propagator is given in Eq.~\eqref{pairproprule}.
}
\label{fig:proppair}
\end{figure}

In XEFT, the charm quark and antiquark numbers $N_c$ and $N_{\bar c}$ are conserved.
In Galilean-invariant XEFT, the pion number $N_\pi$ is also conserved.
These conservation laws can be  built into the Feynman rules 
by appropriate notation for the propagators.  We use a dashed line 
for the pion propagator, a solid line for the $D$ and $\bar D$ propagators,
and a double line consisting of a solid and a dashed line
for the $D^*$ and $\bar D^*$ propagators. 
In the propagators for $D$ and $D^*$, the solid line has a forward arrow. 
In the propagators for $\bar D$ and $\bar D^*$, the solid line has a backward arrow.
The propagators for $D^0$ and $\pi^0$ are illustrated in Fig.~\ref{fig:propDpi}.
The propagator for $D^{*0}$ is illustrated in Fig.~\ref{fig:propD*}.
In XEFT, conservation of charm-quark number requires
 the numbers of solid lines with forward arrows entering and leaving a vertex to be equal.
Conservation of charm-antiquark number requires the numbers of solid lines with backward arrows
entering and leaving a vertex to be equal.
In Galilean-invariant XEFT,  conservation of pion number requires
the numbers of dashed lines entering and leaving a vertex to also be equal.
If the arrows on internal lines of a diagram are omitted, there is an implied sum 
over the two directions of the omitted arrows.
In Galilean-invariant XEFT with a pair field,
the pair propagator is a triple line consisting of two solid lines  and a dashed line,
as illustrated in Fig.~\ref{fig:proppair}.
For simplicity, we omit the opposite arrows on the two solid lines.

The Feynman rules for the propagators of $D^0$ or $\bar D^0$ and $\pi^0$  are
\begin{subequations}
\begin{eqnarray}
\boxed{\frac{i}{E - p^2/(2M) + i \epsilon},}
\label{propD}
\\
\boxed{\frac{i}{E - p^2/(2m) + i \epsilon},}
\label{proppi}
\end{eqnarray}
\label{propDpi}%
\end{subequations}
where $E$ is the kinetic energy of the particle and $\bm{p}$ is its momentum. 
The Feynman rule for the propagator of $D^{*0}$ or $\bar D^{*0}$ with vector indices $i$ and $j$ is
\beq
\boxed{
\frac{i \, \delta^{ij}}{E -  p^2/(2(M+m))  - E_*}, }
\label{propD*}
\eeq
where $E$  is the energy of $D^{*0}$ relative to the $D^0\pi^0$ threshold, $\bm{p}$ is its momentum,
and $E_*$ is the  complex rest energy of  $D^{*0}$ in Eq.~\eqref{E*}.
Galilean invariance requires the kinetic mass of  $D^{*0}$ to be the sum $M+m$ of the kinetic masses of $D^0$ and $\pi^0$.
Since $E_*$ has a negative imaginary part,
an explicit $i \epsilon$ prescription is unnecessary in the $D^{*0}$ propagator in Eq.~\eqref{propD*}.

\begin{figure}[tb]
\centerline{\includegraphics*[
width=7cm,angle=0,clip=true]{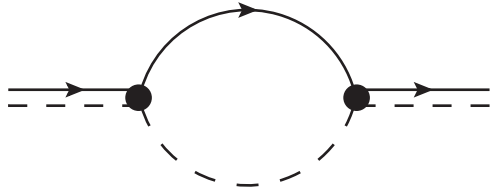}}
\vspace*{0.0cm}
\caption{
The 1-loop $D^{*0}$ self-energy diagram.
This is the only $D^{*0}$ self-energy diagram in Galilean-invariant XEFT.
The Feynman rule for the subdiagram  with the external legs amputated 
is given in Eq.~\eqref{SigmaD*}.
}
\label{fig:D*selfenergy}
\end{figure}

In XEFT beyond LO, there are corrections to the $D^{*0}$ propagator from interactions involving pions.
The $D^{*0}$ self energy $\Sigma(E,p)$ is a function of its energy $E$ and its momentum $p$.
In Galilean-invariant XEFT, the conservation of pion number 
implies that the exact $D^{*0}$ propagator can be calculated analytically
by summing a geometric series in the 1-loop $D^{*0}$ self-energy diagram in Fig.~\ref{fig:D*selfenergy}.
It is useful to have a Feynman rule for this subdiagram.
Galilean invariance implies that the self-energy $ \Sigma$
depends on $E$ and $p$ only through the  invariant energy
\beq	
E_{\rm rest} = E - \frac{p^2}{2(M+m)},
\label{ED*Gal}
\eeq
which is equal to the $D^{*0}$ energy in its  rest frame.
With dimensional regularization in $d$ spatial dimensions,
the Feynman rule for the subdiagram in Fig.~\ref{fig:D*selfenergy}
with the $D^{*0}$ legs amputated is
\beq
\boxed{-i \, \Sigma(E_\mathrm{rest}) \, \delta^{ij} =
i \frac{4 r^2  g_\pi^2}{d} E_\mathrm{rest} \, I_1(E_\mathrm{rest})\, \delta^{ij}, } 
\label{SigmaD*}
\eeq
where $g_\pi^2$ is given in Eq.~\eqref{gpi2} and $r$ is the reduced-mass ratio in Eq.~\eqref{r-mu}.
The function $I_1(E)$ is given by a 1-loop momentum integral:
\beq
I_1(E) =
 \frac{\Gamma(1-d/2)}{(4 \pi)^{d/2}}
\Lambda^{3-d} \left[ e^{- i \pi} 2 \mu_\pi E \right]^{d/2-1} ,
\label{I1-E}
\eeq
where $\Lambda$ is a renormalization scale with dimensions of momentum
and $\mu_\pi$ is the $D^0 \pi^0$ reduced mass in Eq.~\eqref{mupi}.
This loop  integral has a linear ultraviolet divergence that is manifested as a pole in $d-2$.
The analytic function $I_1(E)$ has a branch cut along the positive  real $E$ axis.
The complex phase inside the square brackets in Eq.~\eqref{I1-E} is 
chosen so that it gives the correct branch of the function when $E$ is near $\delta + i \epsilon$ 
on the first sheet and near $E_*$ on the second sheet.

\begin{figure}[tb]
\centerline{\includegraphics*[
width=5cm,angle=0,clip=true]{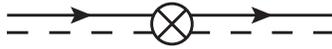}}
\vspace*{0.0cm}
\caption{
The $D^{*0}$ propagator counterterm vertex. Its Feynman rule is given in Eq.~\eqref{D*counterterm}.
}
\label{fig:D*counter}
\end{figure}

The $D^{*0}$ self-energy diagram in Fig.~\ref{fig:D*selfenergy} has linear ultraviolet divergences.
These divergences are cancelled by the $D^{*0}$ propagator counterterm vertex, 
which is illustrated in Fig.~\ref{fig:D*counter}.
A renormalization prescription for the $D^{*0}$  propagator can be expressed in terms of subtractions
to the $D^{*0}$ self-energy.
In the complex on-shell (COS) renormalization scheme 
for the $D^{*0}$ propagator, its pole in $E_\mathrm{rest}$ is at the complex
value $E_*$ in Eq.~\eqref{E*} and the residue of that pole is the same as at LO.
This requires the first two terms in the expansion of $\Sigma(E_{\rm rest})$ in powers of $E_{\rm rest}-E_*$ to be subtracted.
The Feynman rule  for the $D^{*0}$  propagator  counterterm vertex in the COS scheme is
\beq	
\boxed{i \big[ \Sigma(E_*) 
+ \Sigma'(E_*) \, ( E_\mathrm{rest} - E_* )\big]\, \delta^{ij}=
-i \frac{4r^2g_\pi^2}{d} I_1(E_*) \left[ E_* + \frac{d}{2}( E_\mathrm{rest} - E_*) \right]
\delta^{ij}.}
\label{D*counterterm}
\eeq

The Feynman rules for T-matrix elements include external-line factors for incoming and outgoing particles.
For $D^0$, $\bar D^0$ and $\pi^0$, the external-line factors are simply 1.
The corrections to the $D^{*0}$ propagator from interactions involving pions
change the residue of the pole in the $D^{*0}$ propagator by a multiplicative factor $Z_*$.
The external-line factor for $D^{*0}$ includes a residue factor $\sqrt{Z_*}$.
In the COS scheme for the $D^{*0}$ propagator, $Z_*=1$.
The external-line factor for an incoming $D^{*0}$  or $\bar D^{*0}$ 
with polarization vector $\bm{\varepsilon}$ and vector index $i$ is
\beq
\boxed{ \varepsilon^i. }
\label{D*external}
\eeq
The external-line factor for an outgoing $D^{*0}$  or $\bar D^{*0}$ is the same except that
$\varepsilon^i$ is replaced by its complex conjugate.

\subsection{Pair propagator}
\label{sec:LOVertex}

If we use dimensional regularization with $d$ spatial dimensions,
it is useful to introduce a renormalization scale $\Lambda$
to keep the dimensions of coupling constants the same as in the physical dimension $d=3$. 
This would have the effect of multiplying an interaction vertex
with $n$ external lines by $n-2$ powers of $\Lambda^{(3-d)/2}$.
In a Green function with $n$ external legs, the net effect of these powers of  $\Lambda$ is 
a factor of $\Lambda^{3-d}$ for every loop integral
and an overall multiplicative factor of $\Lambda^{(3-d)(n-2)/2}$.
The factor of $\Lambda^{3-d}$ 
associated with a loop integral can be absorbed into its integration measure.
If the Green function is made finite by renormalization,
the overall multiplicative  factor of $\Lambda^{(3-d)(n-2)/2}$
can  simply be discarded, because it is equal to 1 in the physical dimension $d=3$. 
The resulting Feynman rules have no powers of $\Lambda$ in the vertices
and the coupling constants have the same dimensions as in $d=3$.

In XEFT at LO, the only interaction term in the Lagrangian is the contact interaction term in Eq.~\eqref{L-LOint}
for $D^{*0} \bar D^0$ and $D^0 \bar D^{*0}$  in the $C=+$ channel.  
The Feynman rule  is the same for the four vertices for 
$D^{*0} \bar D^0,D ^0 \bar D^{*0} \to D^{*0} \bar D^0,D ^0 \bar D^{*0}$:
$-i \big(1/\sqrt{2}\,  \big)^2 C_0  \delta^{ij}$.
The factors of $1/\sqrt{2}$ come from projecting a pair of charm mesons onto the $C=+$ channel.
Two $C_0$ interactions can be connected by a $D^{*0} \bar D^0$ loop or a $D ^0 \bar D^{*0}$ loop.
In Galilean-invariant XEFT, the loop integral is a function of the invariant energy of the pair of charm mesons:
\beq	
E_\mathrm{cm} = E - \frac{P^2}{2(2M+m)},
\label{Ecm-D*D}
\eeq
where $E$  is their total energy relative to the $D^{*0} \bar D^0$ threshold and $\bm{P}$ is their total momentum.
This invariant energy is equal to the total energy of the pair of charm mesons in their center-of-momentum (CM) frame.
The sum of the two loop diagrams with two $C_0$ interactions can be expressed as the vertex
multiplied by $-2 \mu C_0\, J_1(E_\mathrm{cm})$,  where the function $J_1(E)$ is defined by a 1-loop integral.
Using dimensional regularization in $d$ spatial dimensions, the loop integral  is
\beq	
J_1(E) = 
 \frac{\Gamma(1-d/2)}{(4 \pi)^{d/2}}
\Lambda^{3-d} \, \big[ 2 \mu (E_* - E )  \big]^{d/2-1} ,
\label{J1-E}
\eeq
where $\Lambda$ is the renormalization scale.
This loop integral has a linear ultraviolet divergence that is manifested as a pole in $d-2$. 

The $C_0$ interaction must be treated nonperturbatively in XEFT.
The set of diagrams consisting of an arbitrary number of 
successive $C_0$ interactions connected by $D ^0 \bar D^{*0}$ or $D^{*0} \bar D^0$ loops
is a geometric series that can be summed to all orders analytically.  
The factor in the resulting amplitude  that depends on the invariant energy is
a function ${\cal A}(E_\mathrm{cm})$ with dimensions 1/(momentum). 
In Ref.~\cite{Braaten:2015tga}, $(2\pi/\mu){\cal A}(E)$ was called the {\it LO transition amplitude}.
In this paper, we refer to ${\cal A}(E)$ as the {\it pair propagator},
because it is equal to the propagator for the pair field $\bm{\phi}$ defined in Eq.~\eqref{Phi}
up to a constant multiplicative  factor.  
If we use dimensional regularization in $d$ spatial dimensions, the pair propagator  is
\beq
{\cal A}(E)  = 
\frac{1}{-2 \pi/(\mu C_0) - 4 \pi J_1(E)}.
\label{A0-E:d}
\eeq
Since $ J_1(E)$ has a pole in $d-2$, this amplitude has a finite limit as $d \to 2$ 
only if $1/C_0$  also has a pole in $d-2$.
The coupling constant $C_0$ can be tuned as a function of $d$ so that the LO transition amplitude 
has a finite limit as $d \to 3$:
\beq
{\cal A}(E)  = 
\frac{1}{- \gamma + \sqrt{- 2 \mu (E - E_*)}} \qquad (d=3),
\label{A0-E:3}
\eeq
where $\gamma$ is an interaction parameter that we refer to as the {\it LO binding momentum}.
A  minimal choice for the dependence of $1/C_0$ on $d$ is that it  is the sum of a pole in $d-2$
and a constant:
\beq
\frac{2 \pi}{\mu C_0}   = \left( \frac{2}{d-2} - 2 \right)\Lambda + \gamma.
\label{C0-d}
\eeq
In the limit $d \to 3$, $C_0$ approaches the finite value $2 \pi/(\mu \gamma)$.

The pair propagator in Eq.~\eqref{A0-E:3} has a pole in the energy $E$
at the complex energy
\beq
E_\mathrm{pole,LO}    = E_* - \gamma^2/(2\mu).
\label{EpoleLO}
\eeq
This LO pole energy has an imaginary part $-i \Gamma_{*0}/2$ from the $D^{*0}$ rest energy $E_*$
in Eq.~\eqref{E*}.  This imaginary part gives a contribution to the width of $X$
that can be interpreted as arising
from its decays into $D^0 \bar D^0 \pi^0$ and $D^0 \bar D^0 \gamma$,
which can proceed through the decay of a constituent $D^{*0}$ or $\bar D^{*0}$.
But $X$ also has short-distance decay modes that do not involve decay of a constituent,
such as $J/\psi\, \pi^+\pi^-$.  The contribution to the imaginary part of the pole energy
from these decays can only be taken into account through the imaginary part of $\gamma$ in Eq.~\eqref{EpoleLO}.
It is therefore convenient to take the LO binding momentum $\gamma$ to be a complex interaction parameter.
Unitarity requires the positivity of $\mathrm{Im}[{\cal A}(E)]$  for real energy $E$,
which requires $\mathrm{Im}[\gamma]$ to be positive.
The real and imaginary parts of $\gamma$ are the only interaction parameters in XEFT at LO.

\begin{figure}[tb]
\centerline{\includegraphics*[width=12cm,angle=0,clip=true]{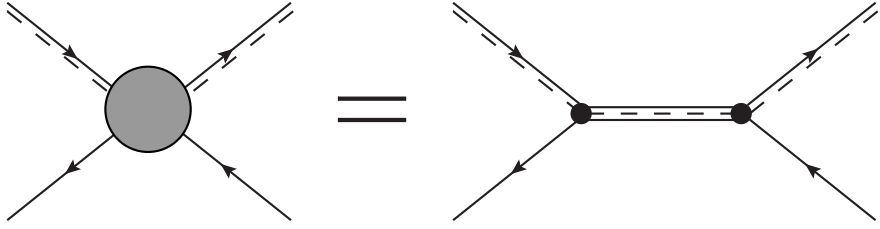}}
\vspace*{0.0cm}
\caption{
The  $D^{*0} \bar D^0$-to-$D^{*0} \bar D^0$ vertex in Ref.~\cite{Braaten:2015tga}
can be expressed as the product of  a pair propagator and two vertices that connect the pair propagator 
to $D^{*0} \bar D^0$ legs.
}
\label{fig:vertex-pairprop}
\end{figure}

\begin{figure}[tb]
\centerline{\includegraphics*[width=12cm,angle=0,clip=true]{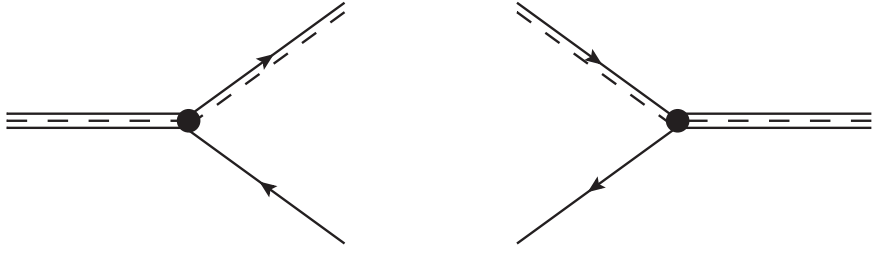}}
\vspace*{0.0cm}
\caption{
Vertices connecting $D^{*0} \bar D^0$ lines to a pair propagator.
Their Feynman rules are given in Eq.~\eqref{DDtoPair}.}
\label{fig:DDtoPair}
\end{figure}

The vertex corresponding  to the LO transition amplitude is the same for the  four transitions
$D^{*0} \bar D^0,D ^0 \bar D^{*0} \to D^{*0} \bar D^0,D ^0 \bar D^{*0}$.
In Ref.~\cite{Braaten:2015tga}, the vertex was represented by a blob, 
as in the diagram on the left side of Fig.~\ref{fig:vertex-pairprop}. 
The Feynman rule for each of the four vertices is
\beq
+i \left(\frac{1}{\sqrt2} \right)^2 
\frac{2 \pi}{\mu}\mathcal{A}(E_\mathrm{cm})\, \delta^{ij}.
\label{amprule}
\eeq
It is convenient to express the $2 \to 2$ vertex whose Feynman rule is given in Eq.~\eqref{amprule}
as the product of a $2 \to 1$ vertex, the pair propagator, and a $1 \to 2$ vertex,
as illustrated in Fig.~\ref{fig:vertex-pairprop}.
The Feynman rule for the propagator of a pair with  energy $E$, momentum $\bm{P}$,
and vector indices $i$ and $j$ is
\beq
\boxed{
-i \frac{2\pi}{\mu}\mathcal{A}(E_\mathrm{cm})\, \delta^{ij},}
\label{pairproprule}
\eeq
where $E_\mathrm{cm}$ is the Galilean-invariant combination of $E$ and $P$ in Eq.~\eqref{Ecm-D*D}.
The vertices connecting $D^{*0} \bar D^0$ lines or  $D ^0 \bar D^{*0}$ lines to a pair propagator are
\beq
\boxed{
-i \frac{1}{\sqrt2}  \, \delta^{ij}.}
\label{DDtoPair}
\eeq
The factor of $1/\sqrt2$ is the amplitude for the pair of charm mesons to be in the $C=+$ channel. 
The vertices connecting $D^{*0} \bar D^0$ lines to a pair propagator are illustrated in Fig.~\ref{fig:DDtoPair}.
It is easy to verify that the product of the pair propagator in Eq.~\eqref{pairproprule}
and two of the vertices in Eq.~\eqref{DDtoPair} reproduces the $2 \to 2$ interaction vertex in Eq.~\eqref{amprule}.

The pair field defined in Eq.~\eqref{Phi} can serve as an interpolating field for the $X$.
The complete pair propagator that takes into account interactions beyond LO
has the same form as in Eq.~\eqref{pairproprule} but with $\mathcal{A}(E_\mathrm{cm})$
replaced by a more complicated function of the energy.
An incoming or outgoing $X$ in a Feynman diagram can be represented by a triple line as in Fig.~\ref{fig:proppair}.
The external-line factor for an incoming $X$ with polarization vector $\bm{\varepsilon}$ and vector index $i$ is
\beq
\boxed{
\sqrt{Z_X} \, \varepsilon^i, }
\label{Xexternal}
\eeq
where $Z_X$ is  the residue of the pole in $E_\mathrm{cm}$ 
of the coefficient of $i \, \delta^{ij}$ in the complete pair propagator analogous to Eq.~\eqref{pairproprule}.
The external-line factor for an outgoing $X$ is the same except that
$\varepsilon^i$ is replaced by its complex conjugate.
The LO pole energy  is the pole in the pair propagator $\mathcal{A}(E)$ in  Eq.~\eqref{A0-E:d}:
\beq
E_\mathrm{pole} = E_* - \frac{2\pi \, \Lambda^2}{\mu} 
\left( \frac{-2\pi/\mu}{\Gamma(1-d/2)\, \Lambda \, C_0}  \right)^{2/(d-2)}.
\label{EpoleLO:d}
\eeq
If $C_0$ is replaced by the expression given by Eq.~\eqref{C0-d},
the LO pole energy reduces to  Eq.~\eqref{EpoleLO} in the limit $d\to 3$.
At LO, $Z_X$ is the residue of the pole in $E$ of $-(2 \pi/\mu) \mathcal{A}(E)$:
\beq
Z_{X,\mathrm{LO}} =  \frac{4\pi^2 \, \Lambda}{\Gamma(2-d/2)\, \mu^2} 
\left( \frac{-2\pi/\mu}{\Gamma(1-d/2)\, \Lambda\,  C_0}  \right)^{(4-d)/(d-2)}.
\label{ZXLO}
\eeq
In the limit $d\to 3$, the LO residue factor reduces to 
\beq
Z_{X,\mathrm{LO}}    = 2 \pi \gamma/\mu^2  \qquad (d=3).
\label{ZXLOd=3}
\eeq
%

\subsection{Pion interaction vertices}
\label{sec:PiVertex}

\begin{figure}[tb]
\centerline{\includegraphics*[width=12cm,angle=0,clip=true]{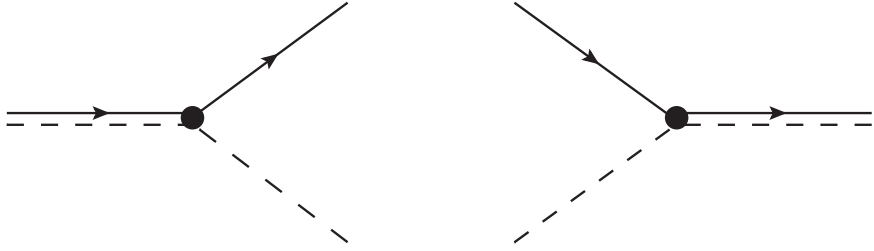}}
\vspace*{0.0cm}
\caption{
The vertices for the pionic transitions $D^{*0} \to D^0\pi^0$ and $D^0\pi^0 \to D^{*0}$. 
The Feynman rules for the vertices are given in Eq.~\eqref{vertex:D*Dpi}.
}
\label{fig:vertpi}
\end{figure}

The charm mesons interact with pions in XEFT through the transitions 
$D^* \leftrightarrow D \pi$ and $\bar D^* \leftrightarrow \bar D \pi$.
The vertices connecting $D^{*0}$ to $D^0\pi^0$ are illustrated in Fig.~\ref{fig:vertpi}.
The Feynman rules for the
$D^{*0} \leftrightarrow D^0\pi^0$ and $\bar D^{*0} \leftrightarrow \bar D^0\pi^0$ vertices 
in Galilean-invariant XEFT are
\beq	
\boxed{ \pm\frac{g}{2 \sqrt{m} f_\pi}
\frac{(M \bm{q} - m\bm{p}_0)^i}{M+m},}
\label{vertex:D*Dpi}
\eeq
where $\bm{q}$ and  $\bm{p}_0$ are the momenta of 
$\pi^0$ and $D^0$ or $\bar D^0$, respectively.
The overall sign is $+$ if the $D^0\pi^0$ or $\bar D^0 \pi^0$ lines are outgoing and $-$ if they are incoming.
In the $D^0 \pi^0$ CM frame defined by $\bm{p}_0 + \bm{q} = 0$, 
the momentum-dependent factor in Eq.~\eqref{vertex:D*Dpi} reduces to $q^i$.
This is the momentum-dependent factor in all frames in original XEFT.

\subsection{NLO interaction vertices}
\label{sec:NewVertices}

In this subsection, we present Feynman rules for NLO interaction vertices in Galilean-invariant XEFT
whose legs include a pair propagator.
The Feynman rules are given below as boxed expressions.
They replace the Feynman rules for NLO interaction vertices in Ref.~\cite{Braaten:2015tga}.

\begin{figure}[tb]
\centerline{\includegraphics*[width=12cm,angle=0,clip=true]{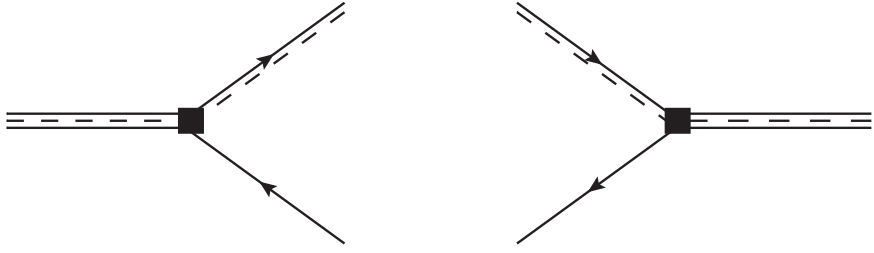}}
\vspace*{0.0cm}
\caption{
The $\nabla^2$ interaction vertices connecting $D^{*0} \bar D^0$ lines to a pair propagator.
The Feynman rules for the vertices are given in Eq.~\eqref{vertex:grad2to1}.
}
\label{fig:NLOvertex2to1}
\end{figure}

The NLO $\nabla^2$  interaction terms in the Lagrangian are given in Eq.~\eqref{Lnabla2}.
The interaction vertices that connect $D^{*0} \bar D^0$ lines to a pair propagator are illustrated in
Fig.~\ref{fig:NLOvertex2to1}.
The Feynman rules for the corresponding vertices 
connecting $D^{*0} \bar D^0$ or $D^0 \bar D^{*0}$ lines to a pair propagator  are
\beq	
\boxed{-i \frac{C_2}{2 \sqrt{2} \, C_0}
\frac{( (M+m) \bm{p}_0 - M \bm{p}_1)^2}{(2M+m)^2} \, \delta^{ij},}
\label{vertex:grad2to1}
\eeq
where $\bm{p}_0$ and $\bm{p}_1$ are the momenta of the spin-0 and spin-1 charm mesons.
These four vertices replace the four $\nabla^2$  vertices in Ref.~\cite{Braaten:2015tga} 
that connect $D^{*0} \bar D^0$ or $D^0 \bar D^{*0}$ lines  to $D^{*0} \bar D^0$ or $D^0 \bar D^{*0}$ lines.
In the CM frame defined by $\bm{p}_0  + \bm{p}_1 = 0$,
the momentum-dependent factor in Eq.~\eqref{vertex:grad2to1} reduces to $p_0^2$.

\begin{figure}[tb]
\centerline{\includegraphics*[width=6cm,angle=0,clip=true]{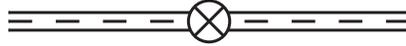}}
\vspace*{0.0cm}
\caption{
The  pair-propagator counterterm vertex.
The Feynman rule for the vertex is given in Eq.~\eqref{vertex:pairpropcount}.
}
\label{fig:NLOcounter}
\end{figure}

The NLO counterterm in the Lagrangian is given in Eq.~\eqref{Lct}.
The pair-propagator counterterm vertex is  illustrated in Fig.~\ref{fig:NLOcounter}.
Its Feynman rule   is
\beq	
\boxed{ -i\frac{1}{C_0^2} \big[ \delta C_0 +D_0 \, (E_\mathrm{cm}-E_*)\big]  \, \delta^{ij}.}
\label{vertex:pairpropcount}
\eeq
This  vertex replaces the  four counterterm vertices in Ref.~\cite{Braaten:2015tga}
that connect $D^{*0} \bar D^0$ or $D^0 \bar D^{*0}$ lines  to $D^{*0} \bar D^0$ or $D^0 \bar D^{*0}$ lines.

\begin{figure}[tb]
\centerline{\includegraphics*[width=12cm,angle=0,clip=true]{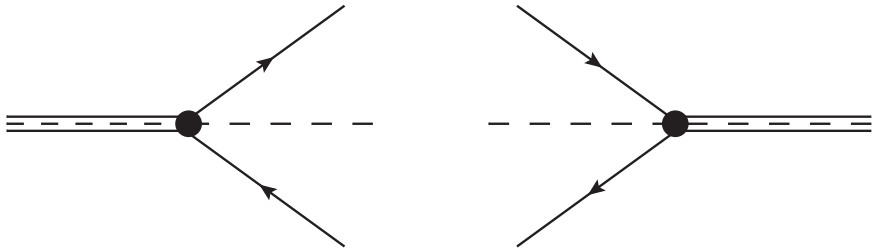}}
\vspace*{0.0cm}
\caption{
The $D \bar D \pi$ interaction vertices connecting $D^0 \bar D^0 \pi^0$ lines to a pair propagator.
The Feynman rules for the vertices are given in Eq.~\eqref{vertex:prop-DDpi}.
}
\label{fig:NLOvertexDDpi}
\end{figure}

The vertices connecting $D^0 \bar D^0 \pi^0$ lines to a pair propagator are illustrated in Fig.~\ref{fig:NLOvertexDDpi}.
The $D \bar D \pi$ interaction terms in the Lagrangian are given in Eq.~\eqref{L-DDpi}.
In Galilean-invariant XEFT, the Feynman rules for the  
vertices connecting  a pair propagator to $D^0 \bar D^0 \pi^0$ lines are
\beq	
\boxed{\pm  \frac{B_1}{\sqrt{2m} \, C_0}
\frac{( 2M \bm{q} - m (\bm{p}_0 + \bm{p}_0^\prime))^i}{2M+m}, }
\label{vertex:prop-DDpi}
\eeq
where $\bm{q}$,  
$\bm{p}_0$, and $\bm{p}_0^\prime$ are the momenta of 
$\pi^0$, $D^0$, and $\bar D^0$, respectively.
The overall sign is $+$ if  the $D^0 \bar D^0\pi^0$ lines are outgoing and $-$ if they are incoming.
In the CM frame defined by $\bm{p}_0  + \bm{p}_0^\prime  + \bm{q} = 0$,
the momentum-dependent factor in Eq.~\eqref{vertex:prop-DDpi} reduces to $q^i$.

\begin{figure}[tb]
\centerline{\includegraphics*[width=12cm,angle=0,clip=true]{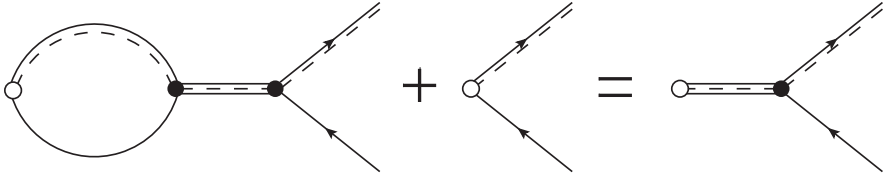}}
\vspace*{0.0cm}
\caption{
A diagrammatic identity that can be used to show that the Feynman rules in Ref.~\cite{Braaten:2015tga}
give the same Green functions as the Feynman rules in this Section.
The open circle on the left side is an arbitrary vertex that connects to $D^{*0} \bar D^0$ or $D^0 \bar D^{*0}$ lines.
The open circle on the right side is the corresponding vertex that connects to a pair propagator.
}
\label{fig:diagid}
\end{figure}

Calculations beyond LO using the  Feynman rules for NLO interaction vertices presented above
are simpler than calculations using the Feynman rules in Ref.~\cite{Braaten:2015tga}.
There are sets of diagrams whose sums are the same with either set of Feynman rules. 
The Feynman rules above give the terms in such a sum more directly and with fewer UV divergences.
The same terms can be obtained using the
Feynman rules in Ref.~\cite{Braaten:2015tga} by using diagrammatic identities.
An example of such a diagrammatic identity is shown in Fig.~\ref{fig:diagid}.
The analytic expression for the identity is
\beq
J_1(E) \, \mathcal{A}(E) + \frac{1}{4 \pi } = -\frac{1}{2\mu C_0} \mathcal{A}(E) .
\label{identity}
\eeq
This identity follows from the expression for the LO transition amplitude in $d$ dimensions in Eq.~\eqref{A0-E:d}.
It allows the loop integral $J_1(E)$ to be cancelled against the loop integral 
in the denominator of the  pair propagator $\mathcal{A}(E)$.

\section{NLO Pair Propagator}
\label{sec:PairPropNLO}

In this Section, we use Galilean-invariant XEFT to calculate the complete pair propagator  to NLO.

\subsection{Complete pair propagator}
\label{sec:CompletePairProp}

The pair propagator in XEFT at LO is given by the Feynman rule in Eq.~\eqref{pairproprule},
where the function $\mathcal{A}(E)$ in $d$ spatial dimensions is given in Eq.~\eqref{A0-E:d}. 
In 3 dimensions, $\mathcal{A}(E)$ reduces to Eq.~\eqref{A0-E:3}.
In XEFT beyond LO, there are corrections to the pair  propagator from interactions involving pions 
and from other interactions.
The corrections to the pair propagator can be organized into a geometric series of pair self-energy diagrams.
The complete pair propagator is obtained by summing the geometric series.
It can be expressed in the form
\beq
-i \frac{2\pi}{\mu}\frac{1}{\mathcal{A}(E_\mathrm{cm})^{-1} - \Pi_0(E_\mathrm{cm})}\, \delta^{ij},
\label{completepairprop}
\eeq
where $ \Pi_0(E)$ is a function with dimensions of momentum that we call the {\it pair self-energy}.  

\begin{figure}[tb]
\centerline{\includegraphics*[width=12cm,angle=0,clip=true]{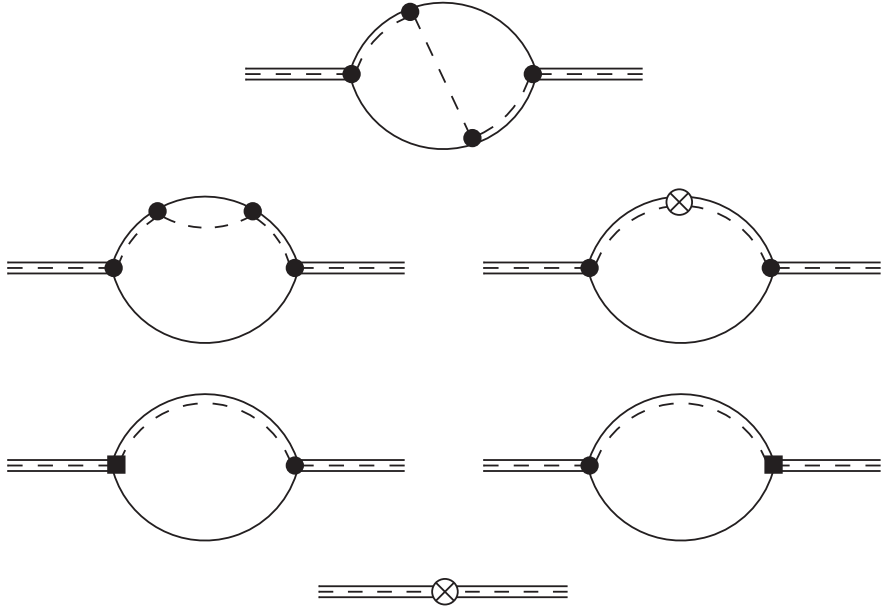}}
\vspace*{0.0cm}
\caption{
The NLO diagrams for the pair self-energy. 
The sum of the amputated diagrams is $i[\mu/(2\pi)]\, \Pi_0(E) \, \delta^{ij}$.
}
\label{fig:NLOpairselfE}
\end{figure}

The NLO diagrams for the pair self-energy are shown in Fig.~\ref{fig:NLOpairselfE}.
The expressions for those diagrams can be obtained from Appendix~\ref{sec:NLOAmp},
in which the NLO diagrams for the transition amplitude for $D^{*0} \bar D^0 \to D^{*0} \bar D^0$
are calculated.  Some of the diagrams for $D^{*0} \bar D^0 \to D^{*0} \bar D^0$
have pair self-energy subdiagrams.  
The expression for each pair self-energy  subdiagram can be obtained by
removing the initial and final vertex factors and the two pair propagators
from the diagram for $D^{*0} \bar D^0 \to D^{*0} \bar D^0$.
The NLO pair self-energy can be expressed as
\beq
\Pi_0(E) = g_\pi^2 \, F(E) + (C_2/C_0)H(E)
 - (2\pi/(\mu C_0^2)) \big[\delta C_0 + D_0 (E-E_*) \big].
\label{Pi0-NLO}
\eeq
The functions $F(E)$ and $H(E)$ can be expressed in terms of the 1-loop integrals $J_n$ and $I_n$
and the 2-loop integrals $K_{lmn}$ defined in Appendix~\ref{sec:LoopInt}.
The function $F(E)$, which has dimensions of (momentum)$^2$, 
comes from the first two rows of diagrams in  Fig.~\ref{fig:NLOpairselfE}.  It can be
obtained from the 2-loop pion-exchange diagram in Eq.~\eqref{A4} and
the  $D^*$ propagator correction diagrams in Eqs.~\eqref{B1} and \eqref{B2}:
\bea
F(E) &=&  
- \frac{8\pi r}{d}
\bigg( \frac{1}{\sqrt{1-r}} 
\big[ 2 K_{110}(E) -2 (2  \mu E_* - r  \mu E) K_{111}(E)  
- (2-r) J_1(E)^2 \big]
\nonumber \\ 
&& \hspace{0.5cm}
+ 2 K_{110}(E) - 4 \mu E_* \big[K_{120}(E) - r  I_1(E_*) J_2(E) \big]
 -  d r  I_1(E_*) J_1(E)   \bigg),
\label{F-E}
\eea
where $r = \mu_\pi/\mu$ is the ratio of reduced masses defined in Eq.~\eqref{r-mu}.
The function $H(E)$, which has dimensions of (momentum)$^3$, comes from 
the third row of diagrams in  Fig.~\ref{fig:NLOpairselfE}.  It can be obtained from the 1-loop
$\nabla^2$ vertex diagrams in Eq.~\eqref{C1234-new}:
\bea
H(E) &=&  
8\pi \mu (E - E_*) \, J_1(E) .
\label{H-E}
\eea
The $\delta C_0$ and  $D_0$  terms in Eq.~\eqref{Pi0-NLO} come from the 
last diagram in  Fig.~\ref{fig:NLOpairselfE}. They can be obtained from the 
pair-propagator counterterm diagram in Eq.~\eqref{D-new}.

\subsection{Renormalization}
\label{sec:Renorm}

The renormalizability of XEFT as an effective field theory requires that ultraviolet (UV) divergences can be cancelled  
order by order in the power counting by renormalization of the  parameters of XEFT.
With dimensional regularization, the UV divergences in loop integrals
produce poles in $d-2$ and poles in $d-3$.  
A pole in $d-3$ represents a logarithmic UV divergence,
and a pole in $d-2$ represents a linear UV divergence.
In the previous calculations in XEFT at NLO in Refs.~\cite{Fleming:2007rp,Jansen:2013cba,Dai:2019hrf},
power divergence subtraction was used to remove the poles in $d-2$.
The subsequent limit $d \to 3$ produces terms 
with positive integer powers of the renormalization scale $\Lambda$.  
It may also produce terms that depend logarithmically on $\Lambda$.
If power divergence subtraction is used to make linear UV divergences explicit in the limit $d \to 3$,
renormalization must remove both the poles in $d-3$ and the  dependence on $\Lambda$.

The Green functions in XEFT should be multiplicatively renormalizable.
A renormalized amputated connected Green function can be defined by multiplying
the amputated connected Green function by an appropriate renormalization factor $\sqrt{Z}$ for every external leg.
For $D^0$ and $\bar D^0$ legs, the renormalization constant is $Z=1$.
For $D^{*0}$ and $\bar D^{*0}$ legs, the renormalization constant 
in the COS scheme for the $D^{*0}$ propagator is $Z_*=1$.
When the Lagrangian for XEFT is formulated using the pair field $\bm{\phi}$ as in Section~\ref{sec:Lagrangian},
there are also Green functions with external pair legs.
We denote the corresponding dimensionless renormalization constant by $Z_\phi$.
It is equal to 1 at LO, but $Z_\phi$ has corrections beyond LO.

The renormalized complete pair propagator differs from the complete pair propagator 
in Eq.~\eqref{completepairprop} by a multiplicative factor $1/Z_\phi$.  
The renormalized pair self-energy $\Pi(E)$ can be defined by
\beq
Z_\phi \big[  \mathcal{A}(E)^{-1} - \Pi_0(E) \big] =  \mathcal{A}(E)^{-1} - \Pi(E).
\label{Pi:renorm}
\eeq
The renormalization constant at NLO can be expressed as  $Z_\phi = 1 + \delta Z_\phi$.
The renormalized  pair self-energy  at NLO is
\bea
\Pi(E)
&=&  g_\pi^2 \, F(E) + (C_2/C_0)\, H(E)
\nonumber\\
&&\hspace{0cm}
- (2\pi/(\mu C_0^2)) \big[\delta C_0 + D_0 \,(E-E_*) \big] - \delta Z_\phi\,  \mathcal{A}(E)^{-1} .
\label{PiNLO}
\eea
It must be possible to choose $\delta Z_\phi$ 
so the linear and logarithmic UV divergences in this expression  all cancel.

We first consider the poles in $d-2$ in the 
renormalized  NLO pair self-energy $\Pi(E)$ in Eq.~\eqref{PiNLO}.
The poles in $d-2$ for the loop integrals are given in subsection~\ref{sec:poles2} of Appendix~\ref{sec:LoopInt}.
The function $F(E)$  in Eq.~\eqref{F-E} has double poles
in $d-2$ from the 2-loop integral $K_{110}$  and from the products of 1-loop integrals 
$J_1^2$ and $I_1 J_1$.  Along with the double poles, which do not depend on $E$,
there are single poles whose coefficients include a logarithm of the form $\log(E_* - E)$.
There are also canceling single poles in $d-2$ in the combination $K_{120}  - r  I_1  J_2$.
The double poles and the constant single poles in $F(E)$
can be cancelled by the counterterm $\delta C_0$ in Eq.~\eqref{PiNLO}.
The function $H(E)$ in Eq.~\eqref{H-E}, which has a factor $E-E_*$,
has a single pole in $d-2$ from the loop integral  $J_1$.
The single pole can be cancelled by the  counterterm $D_0$ in Eq.~\eqref{PiNLO}.
The poles in $d-2$ from the loop integrals in  $\Pi(E)$ that cannot be cancelled by the 
counterterms $\delta C_0$ and $D_0$ are the single poles with energy dependence $\log(E_* - E)$.
The sum of these terms and the $\delta Z_\phi$ term in Eq.~\eqref{PiNLO} are
\beq
\frac{2 r  \sqrt{1-r}\, g_\pi^2 \Lambda^2}{(d-2) \pi}
\log \frac{2 \mu (E_* - E)}{\Lambda^2}   - \delta Z_\phi\,  \mathcal{A}(E)^{-1} .
\label{PiNLO:d->2}
\eeq
The argument of the logarithm has been made dimensionless by using the renormalization scale $\Lambda$.
The expression for the amplitude $\mathcal{A}(E)$ in Eq.~\eqref{A0-E:d} in  the limit $d \to 2$ is
\beq
\lim_{d \to 2} \mathcal{A}(E)  = 
\frac{1/ \Lambda}{\log \big(2 \mu (E_* - E)/\Lambda_2^2 \big)},
\label{A0:d->2}
\eeq
where the momentum scale $\Lambda_2$ in the logarithm is determined by
the constant under the pole in $d-2$ of $C_0^{-1}$.
The dependence on $E$ cancels between the two terms in Eq.~\eqref{PiNLO:d->2}
if $\delta Z_\phi$ has a pole in $d-2$ with the appropriate residue.
The renormalization constant at NLO must have the form
\beq
Z_\phi = 
1 + \left(\frac{2 r  \sqrt{1-r}\, g_\pi^2\,  \Lambda}{(d-2) \pi} + \mathrm{finite} \right),
\label{Zphi:d->2}
\eeq
where the finite NLO term  has a finite limit as $d \to 2$.
The cancellation of the $\log(E-E_*)$ terms leaves a single pole with a factor $\log(\Lambda_2^2/\Lambda^2)$
that can be cancelled by the counterterm $\delta C_0$.
We conclude that all the poles in $d-2$ in the NLO pair self-energy
can be cancelled by the counterterms $\delta C_0$ and $D_0$.

Having verified that all the linear UV divergences in the renormalized pair self-energy
can be cancelled by the counterterms $\delta C_0$ and $D_0$
there is  nothing to be gained by making them explicit using power divergence subtraction.
We therefore choose to simplify intermediate results by using conventional dimensional regularization
in which the only explicit UV divergences are poles in $d-3$.
 
We now consider the poles in $d-3$ in the 
renormalized  NLO pair self-energy $\Pi(E)$ in Eq.~\eqref{PiNLO}.
The poles in $d-3$ for the loop integrals are given in subsection~\ref{sec:poles3} of Appendix~\ref{sec:LoopInt}.
The only poles in $d-3$ come from the 2-loop integrals 
$K_{110}$, $K_{120}$, and $K_{111}$ in the function $F(E)$ in Eq.~\eqref{F-E}.
The poles in $K_{120}$ and $K_{111}$ are constants. 
The pole in $K_{110}$ is a linear function of $E$. 
Thus all the poles in $d-3$ can be cancelled by the counterterms $\delta C_0$ and $D_0$.
We conclude that the logarithmic UV divergences in the  NLO pair self-energy
can be cancelled by  these counterterms.

\subsection{Minimal subtraction renormalization scheme}
\label{sec:RenormMS}

We have verified that the  linear and logarithmic UV divergences in the NLO pair 
self-energy can be cancelled by the counterterms $\delta C_0$ and $D_0$
and the NLO term $\delta Z_\phi$  in the pair renormalization constant.
A renormalization scheme for 
$D^{*0} \bar D^0$ and $D^0 \bar D^{*0}$ amplitudes in XEFT corresponds to a specific choice for those counterterms.
The simplest renormalization scheme is the  {\it minimal subtraction} (MS) scheme,
in which $\delta C_0$, $D_0$, and $\delta Z_\phi$ are chosen to cancel only the poles in $d-3$.
Since $\delta Z_\phi$ has a pole in $d-2$ but no poles in $d-3$, $\delta Z_\phi=0$ in the MS scheme.
At NLO, the poles in $d-3$ in the pair propagator appear only in the function $F(E)$ in Eq.~\eqref{F-E}.
The explicit form of the poles in $d-3$ of $F(E)$  is
\bea
\big[ F(E)\big]_\mathrm{pole}  &=& \frac{r^2}{3\pi^2(d-3)}
\Bigg\{ \left[ \left( \frac{1}{\sqrt{1-r}} + 3\right) \sqrt{r} 
-  \frac{2-r}{1-r} \arccos\big(\sqrt{r}\, \big) \right] \mu E_*
\nonumber\\
&&\hspace{2.5cm} 
- \left[ \left(\frac{1}{\sqrt{1-r}}+ 1\right) \sqrt{r} - \frac{r}{1-r} \arccos\big(\sqrt{r} \,\big) \right] \mu (E - E_*) \bigg\}.~~~
\label{Fpole}
\eea
A pole in $d-3$ of $F(E)$ is accompanied by  the logarithm  $\log(2 \mu E_*/\bar \Lambda^2)$,
where $\bar \Lambda$ is a renormalization scale.
In the MS scheme, the cancellation of the poles in $d-3$ by the counterterms leaves terms in $F(E)$
that depend on  $\bar \Lambda$. 
They can be obtained by replacing $1/(d-3)$ in Eq.~\eqref{Fpole} by $\log(2 \mu E_*/\bar \Lambda^2)$.
The logarithm multiplying the constant $\mu E_*$ term can be absorbed into 
an NLO correction to the LO parameter $\gamma$.
However there is also a logarithm multiplying the $\mu (E- E_*)$ term. 
The renormalization scale $\bar \Lambda$ in this logarithm
can be interpreted as an additional real-valued interaction parameter in the MS scheme
associated with renormalization of the coupling constant $D_0$.
The existence of this additional  interaction parameter was not recognized in Ref.~\cite{Braaten:2015tga}.

\subsection{Complex threshold renormalization scheme}
\label{sec:RenormCT}

We  introduce a  new renormalization scheme for $D^{*0} \bar D^0$ and $D^0 \bar D^{*0}$ amplitudes in XEFT 
that we call the complex  threshold (CT) renormalization scheme.
It is defined by specifying the behavior of the renormalized pair propagator near the complex threshold $E=E_*$.
The renormalized inverse pair propagator has a threshold expansion 
in half-integer powers of $E-E_*$ or, equivalently,
in integer powers of the function $\kappa(E)$ defined by
\beq
\kappa(E) = \sqrt{2 \mu (E_*-E)}.
\label{kappa-E}
\eeq
The CT scheme is partly defined by specifying the first two leading terms in the threshold expansion:
\beq
\mathcal{A}(E)^{-1} - \Pi(E) =  -\gamma + \kappa(E)  + \mathcal{O} \big( \kappa^2(E) \big).
\label{CTdef}
\eeq
At NLO, the definition of the CT scheme is completed by specifying  the real part of the 
coefficient of $\kappa^2(E)$ in the threshold expansion of $\Pi(E)$.  
We choose to denote that real part by $F_2 g_\pi^2$,
where $F_2$ is a dimensionless adjustable interaction parameter.
At higher orders, the definition of the CT scheme may need to be extended by specifying 
the real parts of coefficients of higher  integer powers of $\kappa^2(E)$.

We proceed to obtain a more explicit expression for the renormalized NLO pair self-energy $\Pi(E)$ in the limit $d \to 3$.
The expression for $\Pi(E)$ in Eq.~\eqref{PiNLO} depends on the functions $F(E)$ and $H(E)$,
which are expressed in terms of loop integrals in Eqs.~\eqref {F-E} and \eqref{H-E}.
The threshold expansions in powers of $E-E_*$ of the  loop integrals in Ref.~\cite{Braaten:2015tga}
are given in subsection~\ref{sec:threshexp} of Appendix~\ref{sec:LoopInt}.
In the limit $d \to 3$, the function $H(E)$ is very simple:
\beq
H(E) = \kappa^3(E) .
\label{Hexpand}
\eeq
In the limit $d \to 3$, the function $F(E)$ can be expanded in integer powers of $\kappa(E)$:
\beq
F(E) = f_0 \, \kappa_*^2 + f_1 \, \kappa_* \kappa(E) + f_2 \, \kappa^2(E) + f_4 \, \kappa^4(E)/\kappa_*^2 + \ldots.
\label{Fexpand}
\eeq
where $\kappa_* = \sqrt{2 \mu E_*}$.
The dimensionless coefficients $f_n$ are functions of the reduced-mass ratio $r=\mu_\pi/\mu$.
The $f_1$ term  is the only one with an odd power of $\kappa(E)$.
The coefficient $f_1$ is pure imaginary, and it is suppressed by a factor of $r^{5/2}$:
\beq
f_1 =  i \frac{r^{5/2}}{3 \pi \sqrt{1-r}} .
\label{f1}
\eeq
The coefficients $f_0$ and $f_2$ have single poles in $d-3$,  and they have finite imaginary parts:
\begin{subequations}
\bea
\mathrm{Im}[f_0] &=& 
\frac{r^2}{6\pi} \left[ \frac{2-r}{1-r} \arccos  \big(\sqrt{r}\,\big) - \left( 3 +\frac{1}{\sqrt{1-r}} \right) \sqrt{r}  \right],
\label{Imf0}
\\
\mathrm{Im}[f_2] &=&  - \frac{r^{5/2}}{6 \pi} 
\left[1+\frac{1}{\sqrt{1-r}} - \frac{\sqrt{r}}{1-r} \arccos  \big(\sqrt{r}\, \big)\right]. 
\label{Imf2}
\eea
\end{subequations}
All the higher coefficients  have finite limits as $d \to 3$, and they are real valued.
The coefficients $f_n$ with $n \ge 2$ can be expressed analytically in terms of hypergeometric functions.
The coefficient $f_4$ is
\begin{eqnarray}
f_4 &=& \frac{2(2-r)}{3\pi^2\sqrt{1-r}}
\bigg( {}_2F_1\big(\mathrm{-}\tfrac12, \mathrm{-} \tfrac12,\tfrac32; 1-r \big)
+ \frac{2-4r - r^2}{8 \sqrt{1-r}} \, \arccos \big( \sqrt{r} \big) 
\nonumber\\
&& \hspace{3cm}  
+ \frac{(12 -22r -5 r^2)\sqrt{r}}{24(2-r)}  -  \frac{\sqrt{1-r}  \, r^{5/2}}{40(2-r)} \bigg).
\label{f4-anal} 
\end{eqnarray}
Its value in the limit $r \to 0$ is  $f_4 = 2/(3\pi)$.
Its numerical value at $r=0.1304$   is  $f_4 = 0.1979$. 
The renormalized NLO pair self-energy in Eq.~\eqref{PiNLO} can be expressed as
\bea
\Pi(E) &=& 
 \big[   f_0 g_\pi^2 \kappa_*^2 - 2\pi\delta C_0/(\mu C_0^2) + \delta Z_\phi \gamma \big]
+ \big[   f_1 g_\pi^2 \kappa_* - \delta Z_\phi  \big] \kappa(E) 
\nonumber\\
&&+ \big[ f_2 g_\pi^2 + \pi D_0/(\mu^2 C_0^2) \big] \kappa^2(E)
+ (C_2/C_0)\,\kappa^3(E)
+ g_\pi^2 \,  F_4(E),
\label{PiNLO:renorm}
\eea
where $F_4(E) $ is obtained by subtracting from $F(E)$
the first three  terms in its expansion in powers of $\kappa(E)$:
\beq
F_4(E)  = F(E) -  \big[ f_0\, \kappa_*^2 + f_1 \,   \kappa_* \kappa(E) + f_2 \, \kappa^2(E) \big].
\label{F4-E}
\eeq
It has a threshold expansion in even powers of $\kappa(E)$ that begins with a $\kappa^4(E)$ term.

We proceed to implement the CT scheme for the complete pair propagator at NLO.
The first four terms in the threshold expansion for the NLO renormalized pair self-energy $\Pi(E)$ 
are already explicit in Eq.~\eqref{PiNLO:renorm}.
There are poles in $d-3$ in the coefficients $f_0$ and $f_2$.
The CT scheme requires the total subtraction of the  leading term and the $\kappa(E)$ term in Eq.~\eqref{PiNLO:renorm}
and the partial subtraction of the $\kappa^2(E)$ term.

We first consider the  $\kappa(E)$ term in the threshold expansion for  $\Pi(E)$ in Eq.~\eqref{PiNLO:renorm}.
The total subtraction of the $\kappa(E)$ term  requires 
$\delta Z_\phi = f_1 g_\pi^2 \kappa_*$.  The coefficient $f_1$ is finite at $d=3$ and pure imaginary.
The resulting expression for the renormalization constant $Z_\phi$ in the CT scheme is
\beq
Z_\phi = 1 +  i \frac{r^{5/2}}{3 \pi \sqrt{1-r}} g_\pi^2 \kappa_* \quad (d=3).
\label{ZphiCT}
\eeq
If we include in $\delta Z_\phi$ the pole in $d-2$ in Eq.~\eqref{Zphi:d->2},
we must also include an additional finite term that cancels that pole term at $d=3$,
so that $Z_\phi$ in the limit $d \to 3$ is again given by Eq.~\eqref{ZphiCT}.
Since the finite renormalization in Eq.~\eqref{ZphiCT} is not an essential aspect of the CT scheme,
we will continue to regard $\delta Z_\phi$ as arbitrary.

We next consider the  leading term in the threshold expansion for  $\Pi(E)$ in Eq.~\eqref{PiNLO:renorm}.
The NLO binding momentum with $\delta Z_\phi = f_1 g_\pi^2 \kappa_*$ is
\beq
\gamma_\mathrm{NLO} =  \gamma + f_1 \, g_\pi^2 \kappa_*\gamma + f_0 \, g_\pi^2 \kappa_*^2
 - 2\pi\delta C_0/(\mu C_0^2),
\label{gammaNLO}
\eeq
where $f_1$ is given in Eq.~\eqref{f1} and the imaginary part of $f_0$ is given  in Eq.~\eqref{Imf0}.
The  terms $f_1 \, g_\pi^2 \kappa_*\gamma $ and $i \, \mathrm{Im}[f_0] \, g_\pi^2 \kappa_*^2$ 
give positive  contributions to the imaginary part of $\gamma_\mathrm{NLO}$.
These contributions, which take into account the decay of $X$ into $D^0 \bar D^0 \pi^0$,
 add to the imaginary part of $\gamma$, which takes into account
short-distance decays of $X$,  such as its decay into $J/\psi\, \pi^+\pi^-$.
The counterterm $\delta C_0$ must cancel  the pole  in $d-3$
from the coefficient $f_0$ in Eq.~\eqref{gammaNLO}.
It could be chosen to  cancel also  the real finite part of $f_0$.
The CT scheme requires the total subtraction of the  leading  term in  Eq.~\eqref{PiNLO:renorm},
which implies $\gamma_\mathrm{NLO} =  \gamma$.

We finally consider the $\kappa^2(E)$ term in the threshold expansion for  $\Pi(E)$ in Eq.~\eqref{PiNLO:renorm}.
The counterterm $D_0$ must  cancel the pole in $d-3$ from the coefficient $f_2$ in Eq.~\eqref{PiNLO:renorm}.
It can be chosen to also cancel  an arbitrary real part of $f_2$.
We denote the remaining finite real part of that coefficient  by $F_2$.
The resulting expression for the renormalized pair self-energy at NLO  is
\beq
\Pi(E) =  \big[   f_1 g_\pi^2 \kappa_* - \delta Z_\phi  \big] \kappa(E) 
+ \big( F_2+ i\, \mathrm{Im}[f_2] \big) g_\pi^2 \, \kappa^2(E)
+ (C_2/C_0)\,\kappa^3(E)
+ g_\pi^2 \,  F_4(E).
\label{PiNLO:CT}
\eeq
In the CT scheme, the term proportional to $ \kappa(E)$ is absent.
The adjustable real interactions parameters in XEFT at NLO in the CT scheme are 
 $\mathrm{Re}[\gamma]$, $\mathrm{Im}[\gamma]$, $F_2$, and $C_2/C_0$.
The existence of the additional  interaction parameter $F_2$ was not recognized in Ref.~\cite{Braaten:2015tga}.

\subsection{Other previous renormalization schemes}
\label{sec:RenormPrevkous}

In  Refs.~\cite{Fleming:2007rp,Dai:2019hrf},
the differential decay rate of the $X$ into $D^0 \bar D^0 \pi^0$ was calculated to NLO in original XEFT. 
The  LO binding momentum  $\gamma$ was taken to be a real adjustable parameter.
The calculation of the NLO diagrams did not produce any poles in $d-3$.
Power divergence subtraction was used to make the poles in $d-2$ explicit as
dependence on the renormalization scale $\Lambda$, which was denoted by $\Lambda_\mathrm{PDS}/2$.
The pair propagator $\mathcal{A}(E)$  in $d$ dimensions is given in Eq.~\eqref{A0-E:d}.
If the pole in $d-2$ is subtracted from the loop integral $J_1(E)$ in Eq.~\eqref{J1-E}, 
then the pole in $d-2$ must also be subtracted from $2 \pi/(\mu C_0)$ in Eq.~\eqref{C0-d}.
The resulting expression for  $C_0$ satisfies
\beq
C_0\, (2 \Lambda - \gamma)  = - 2\pi/\mu.
\label{C0-Lambda}
\eeq
In  Refs.~\cite{Fleming:2007rp,Dai:2019hrf}, the remaining dependence on
the renormalization scale appeared in the following combinations:
\begin{subequations}
\begin{eqnarray}
&&
C_2\, (2 \Lambda - \gamma)^2  = (\pi/\mu) \, R_0,
\label{C2:renorm}
\\
&&
\big[ B_1+(g \mu/f_\pi)C_2\big] \, (2 \Lambda - \gamma)  = \eta/(100~\mathrm{MeV})^3.
\label{B1:renorm}
\end{eqnarray}
\label{C2,B1:renorm}%
\end{subequations}
These equations defined two renormalized interaction parameters: 
$R_0$ (denoted by $r_0$ in Ref.~\cite{Fleming:2007rp}), which has dimensions 1/(momentum), 
and $\eta$, which is dimensionless.
In  Refs.~\cite{Fleming:2007rp,Dai:2019hrf},
$R_0$ was assumed to be positive and less than $1/(100\, \mathrm{MeV})$
and $\eta$ was assumed to be in the range $|\eta| < 1$.

In our calculations, we have chosen not to make linear UV divergences manifest
by using power divergence subtraction.
The same results could be obtained by using power divergence subtraction and then setting $\Lambda =0$.
If we divide Eq.~\eqref{C2:renorm} by Eq.~\eqref{C0-Lambda} and then set $\Lambda =0$,
we find
\beq
C_2/C_0  = R_0/(2 \gamma).
\label{C2-r0}
\eeq
Thus we can replace $C_2/C_0$ in our NLO corrections by $R_0/(2 \gamma)$,
where $R_0$  is the renormalized interaction parameter introduced in Ref.~\cite{Fleming:2007rp}.

In Ref.~\cite{Jansen:2013cba}, the $D^{*0} \bar D^0$ scattering length was calculated to NLO  in original XEFT.
In addition to terms proportional to $\Lambda$, $\Lambda^2$, $\log \Lambda$,
there was   explicit dependence on $\Lambda$ from  terms proportional to $1/(2\Lambda-\gamma)$.
These terms  were introduced by a resummation prescription
for dealing with an infrared divergence at the $D^{*0} \bar D^0$ threshold.
The terms proportional to $1/(2\Lambda-\gamma)$  were not removed by  the renormalization of the parameters. 
This failure of the renormalization procedure suggests that the resummation prescription
for the infrared divergences  in Ref.~\cite{Jansen:2013cba} was incompatible with the renormalization prescription.

In Ref.~\cite{Braaten:2015tga}, which introduced Galilean invariant XEFT, a renormalization prescription for 
$D^{*0} \bar D^0$ and $D^0 \bar D^{*0}$ amplitudes 
called the {\it complex on-shell} (COS) renormalization scheme was  introduced.
The COS scheme requires the pole in the energy and the residue of the pole in the 
$D^{*0} \bar D^0$ transition amplitude to be the same as at LO.
This renormalization scheme will be discussed in Section~\ref{sec:RenormCOS} 
after the NLO calculation of the $D^{*0} \bar D^0$ transition amplitude.

\subsection{NLO pole energy}
\label{sec:EpoleNLO}

The complete pair propagator $\Pi(E)$ has a pole in $E$ at a complex energy that is conveniently expressed as
\beq	
E_\mathrm{pole} =  E_*- \gamma_X^2/(2\mu) ,
\label{Epole}
\eeq
where $\gamma_X$ is the complex binding momentum.
The pole energy at NLO in the CT scheme is  a zero of the renormalized inverse pair propagator 
$-\gamma +\kappa(E)-\Pi(E)$, where $\Pi(E)$ is given in Eq.~\eqref{PiNLO:CT}. 
The equation for $\gamma_X$ at NLO can be obtained by  substituting 
$E \to E_\mathrm{pole}$ in the expression for $\Pi(E)$ and then setting it  equal to $-\gamma+\gamma_X$.
The equation can be expressed as
\begin{equation}
\gamma_X =  \gamma + \big[   f_1 g_\pi^2 \kappa_* - \delta Z_\phi  \big]  \gamma_X
+ \big( F_2+ i\,  \mathrm{Im}[f_2]   \big) \, g_\pi^2  \gamma_X^2
+ \frac{R_0}{2\gamma}  \, \gamma_X^3 + g_\pi^2\,  F_4\big(E_*  -\gamma_X^2/(2\mu) \big).
\label{gammaXNLO:eq}
\end{equation}
In the CT scheme, the term on the right side proportional to $\gamma_X$ is zero.
We have used Eq.~\eqref{C2-r0} to set  $C_2/C_0 = R_0/(2\gamma)$.
The equation for $\gamma_X$ can be solved as an expansion in powers of $\gamma$.
The solution for the first few terms in the CT scheme is
\begin{eqnarray}
\gamma_X &=& 
\gamma +  \tfrac12 \big[ R_0 +2 \big( F_2+ i\,  \mathrm{Im}[f_2] \big) g_\pi^2 \big] \gamma^2
\nonumber\\
&& \hspace{0.5cm}
+ \tfrac34\big[ R_0 +2 \big( F_2+ i\,  \mathrm{Im}[f_2] \big) g_\pi^2 \big]
\big[ R_0 + \tfrac43 \big( F_2+ i\,  \mathrm{Im}[f_2] \big) g_\pi^2 \big] \gamma^3
 + \ldots.
\label{gammaXNLO:exp}
\end{eqnarray}

In the calculation of the rate for a reaction with $X$ as an incoming  or outgoing particle,
the T-matrix element has a factor $\sqrt{Z_X}$ for the $X$.
The residue factor  $Z_X$ is determined by the derivative with respect to $E$
of the renormalized inverse pair propagator in Eq.~\eqref{Pi:renorm}  evaluated at $E_\mathrm{pole}$.
The reciprocal of the residue factor can be expressed as
\begin{equation}
Z_X^{-1} =  Z_{X,\mathrm{LO}}^{-1} +\big (\mu/(2 \pi)\big)\,  \Pi^\prime(E_\mathrm{pole}),
\label{ZXNLO} 
\end{equation}
where $Z_{X,\mathrm{LO}}$ is the LO residue factor in Eq.~\eqref{ZXLO}.
The NLO pair self-energy $\Pi(E)$ is given in Eq.~\eqref{PiNLO:CT}.
The reciprocal of the residue factor at NLO in the limit $d \to 3$ is
\begin{eqnarray}
Z_{X,\mathrm{NLO}}^{-1} &=& \frac{\mu^2}{2 \pi}\bigg[
\frac{1}{\gamma} - \big(   f_1 g_\pi^2 \kappa_* - \delta Z_\phi  \big)  \frac{1}{\gamma_X}
-  2\big( F_2+ i\,  \mathrm{Im}[f_2]   \big) \, g_\pi^2 
\nonumber\\
&& \hspace{1cm}
-\frac32 \frac{R_0\, \gamma_X}{\gamma}   + \frac{g_\pi^2}{\mu}  F_4^\prime\big(E_*  -\gamma_X^2/(2\mu) \big) \bigg]
\qquad (d=3).
\label{ZXNLO:d=3}
\end{eqnarray}
Note that the choice $Z_\phi = f_1 g_\pi^2 \kappa_*$ in the CT scheme
simplifies both the equation for the binding momentum in Eq.~\eqref {gammaXNLO:eq} 
and Eq.~\eqref{ZXNLO:d=3} for the residue factor.

\section{$\bm{D^{*0} \bar{D}^0}$ Scattering}
\label{sec:D*Dscat}

In this Section, we use Galilean-invariant XEFT to calculate the $D^{*0} \bar D^0$ elastic scattering amplitude
 to NLO and we discuss the breakdown of the effective range expansion.

\subsection{NLO transition amplitude}
\label{sec:D*DAmp}

The amputated connected Green function for $D^{*0} \bar D^0 \to D^{*0} \bar D^0$
is a tensor $\mathcal{T}^{ij}$ whose vector indices are  those of the incoming and outgoing $D^{*0}$ lines.
If the incoming and outgoing $ \bar D^0$ are on their energy shells
but the incoming and outgoing $D^{*0}$ are off their energy shells,
this transition tensor $\mathcal{T}^{ij}(E, \bm{p},\bm{p}^\prime)$
is a function of the total energy $E$ of the pair of charm mesons in their CM frame
and the relative momenta $\bm{p}$ and $\bm{p}'$ of the incoming and outgoing charm mesons.
The transition tensor  $\mathcal{T}_{+}^{ij}$
for the $C=+$ channel is  the sum of the transition tensors for
$D^{*0} \bar D^0 \to D^{*0} \bar D^0$, $D^{*0} \bar D^0 \to D^0 \bar D^{*0}$,
$D^0 \bar D^{*0} \to D^{*0} \bar D^0$, and $D^0 \bar D^{*0} \to D^0 \bar D^{*0}$ multiplied by 1/2.
The S-wave contribution to the transition tensor
can be obtained by averaging over the directions of $\bm{p}$ and $\bm{p}^\prime$. 
It is diagonal in the vector indices $i$ and $j$, and it is a function of $E$, $p$, and $p^\prime$.
The $C=+$ S-wave transition tensor can be expressed in the form
\beq	
\Big\langle \mathcal{T}_{+}^{ij}(E,\bm{p},\bm{p}') \Big\rangle_{\bm{\hat p},\bm{\hat p}^\prime}  
= \frac{2 \pi}{\mu} \mathcal{A}_{s+}(E,p,p^\prime) \, \delta^{ij},
\label{A-Swave}
\eeq
where the scalar transition amplitude  $\mathcal{A}_{s+}$ has dimensions 1/momentum.
At LO, the $C=+$ S-wave transition amplitude reduces to the pair propagator $\mathcal{A}(E)$  in Eq.~\eqref{A0-E:3}.

The NLO diagrams for the transition tensor
for $D^{*0} \bar D^0 \to D^{*0} \bar D^0$ are calculated in Appendix~\ref{sec:NLOAmp}.
There are  three pion-exchange diagrams shown in Fig.~\ref{fig:NLOpiex}:
 two 1-loop diagrams and a 2-loop diagram.
There are two $D^*$ propagator correction diagrams shown in Fig.~\ref{fig:NLOD*se}:
a 2-loop diagram with a $D^*$ self-energy subdiagram 
and a 1-loop diagram with a $D^*$ self-energy counterterm.
With the Feynman rules for NLO interaction vertices in Section~\ref{sec:NewVertices},
there are four $\nabla^2$ vertex diagrams in Fig.~\ref{fig:NLOnabla2:3}: two tree diagrams and two 1-loop diagrams.
There is also a  tree diagram with a pair-propagator counterterm vertex in Fig.~\ref{fig:NLOcount:3}.
There are analogous diagrams for the other three amplitudes $D^{*0} \bar D^0 \to D^0 \bar D^{*0}$,
$D^0 \bar D^{*0} \to D^{*0} \bar D^0$, and $D^0 \bar D^{*0} \to D^0 \bar D^{*0}$.
The amplitudes for $D^{*0} \bar D^0 \to D^0 \bar D^{*0}$ and $D^0 \bar D^{*0} \to D^{*0} \bar D^0$
also have a tree-level pion exchange diagram.
The tree diagram for $D^{*0} \bar D^0 \to D^0 \bar D^{*0}$ is shown in Fig.~\ref{fig:NLOpiextree}.

The complete NLO $C=+$ S-wave  transition amplitude can be expressed as
\bea
\mathcal{A}_{s+}(E,p,p^\prime) &=& \mathcal{A}(E) + \mathcal{A}_\pi(E,p,p^\prime)
\nonumber \\
&& \hspace{-1cm}
+ \Big( g_\pi^2 \, \big[ G(E,p) + G(E,p^\prime) \big] + (C_2/C_0)(p^2 + p^{\prime 2})/2 \Big)\mathcal{A}(E) 
+  \Pi_0(E)\,  \mathcal{A}^2(E),
\label{As+NLO}
\eea
where $\Pi_0(E)$ is the pair self-energy given in Eq.~\eqref{Pi0-NLO}.
The  tree-level pion-exchange term $\mathcal{A}_\pi$  can be obtained from Eq.~\eqref{Tpi-S}:
\bea
\mathcal{A}_\pi(E,p,p^\prime)  &=& -
 \frac{r g_\pi^2}{6\pi  \sqrt{1-r}} \Bigg( \frac{ 2 (2-r) r \mu E - r (p^2+p^{\prime 2})}{4 \sqrt{1-r} \, p \,p^\prime}
\nonumber\\
&& \hspace{02cm}
\times \log \frac{2 r \mu E - p^2- p^{\prime 2} +  2 \sqrt{1-r}\,  p\,p^\prime}
{2 r \mu E - p^2- p^{\prime 2} - 2 \sqrt{1-r}\,  p \,p^\prime}
- (2-r) \Bigg),
\label{Api-S}
\eea
where $r = \mu_\pi/\mu$. 
The function $G(E,p)$, which has dimensions of momentum, 
comes from the 1-loop pion-exchange diagram in Eq.~\eqref{AS,3}:
\bea
G(E,p) &=&
 \frac{2 r}{d \sqrt{1-r}}  
 \Big( - r\, L_0(E,p) +  r \big[2\mu E_* + 2 (1-r)\mu E   -p^2\big] L_1(E,p) 
 \nonumber\\
 && \hspace{4cm}
 + (2-r) J_1(E) \Big)  .
 \label{G-E,p}
\eea

The terms proportional to $\mathcal{A}(E)$ in Eq.~\eqref{As+NLO} have a single pole in the energy $E$. 
The terms proportional to $\mathcal{A}^2(E)$ in Eq.~\eqref{As+NLO} have an unphysical double pole in $E$.
The N$^2$LO contribution would have a triple pole and higher order contributions would have even higher poles.  
These unphysical multiple poles can be summed to all orders, in which case they 
produce a shift in the position of the single pole in the LO amplitude.
An expression for the $C=+$ S-wave transition amplitude that has NLO accuracy but only a single pole in $E$  is
\beq
\mathcal{A}_{s+}(E,p,p^\prime) = \mathcal{A}_\pi(E,p,p^\prime) 
+ W_{\pi 0}(E,p)\, \frac{1}{ \mathcal{A}(E)^{-1} - \Pi_0(E)}\, W_{\pi 0}(E,p^\prime),
\label{A-NLO:frac}
\eeq
where $\Pi_0(E)$ is the NLO pair self-energy  in Eq.~\eqref{Pi0-NLO}.
The numerator factors at NLO are given by
\beq
W_{\pi 0}(E,p) = 1 + g_\pi^2 \, G(E,p) + (C_2/C_0)\, p^2/2.
\label{N:frac}
\eeq
By expressing the numerator as a product as in Eq.~\eqref{A-NLO:frac}, the residue of the pole at $E = E_\mathrm{pole}$
is guaranteed to factor into the product of a function of the incoming relative momentum $p$ 
and a function of the outgoing relative momentum $p^\prime$.
The denominator factor in Eq.~\eqref{A-NLO:frac} is the complete pair propagator at NLO, which
can be obtained by summing the geometric series of NLO pair self-energy diagrams.  

In Ref.~\cite{Braaten:2015tga}, the NLO $C=+$ S-wave transition  amplitude in Eq.~\eqref{As+NLO} was calculated 
only in the limit $p\to 0$, $p^\prime \to 0$.
The numerator factor $W^2_{\pi 0}(E,0)$ was expanded to NLO and expressed as
 $1 + g_\pi^2 G(E)$, where $G(E)= 2\,  G(E,0)$.
It was stated in Ref.~\cite{Braaten:2015tga} that to NLO accuracy, 
the $g_\pi^2 G(E)$ terms in the numerator could equally well be moved to the denominator
in a factor multiplying $\mathcal{A}(E)^{-1}$.
This is incorrect, because it changes the  pole energy. 
The numerator and the denominator in Eq.~\eqref{A-NLO:frac} could however both  be multiplied
by the same constant $Z_\phi$.

\subsection{Renormalization}
\label{sec:D*DRenorm}

The renormalized transition amplitude for $D^{*0} \bar D^0$ in the $C=+$ S-wave channel
can be obtained from the transition amplitude in Eq.~\eqref{A-NLO:frac}
by multiplying it by the appropriate  renormalization constants for the external $D^{*0}$ lines.
If the COS renormalization is used for the $D^{*0}$ propagator,
the renormalization constant  for  $D^{*0}$ is $Z_*=1$.
Thus the  transition amplitude in Eq.~\eqref{A-NLO:frac} must be UV finite.
There are however UV divergences in the pair propagator $\Pi_0(E)$ and in the numerator factors.
A manifestly finite expression for the transition amplitude 
can be obtained by multiplying the numerator and denominator of Eq.~\eqref{A-NLO:frac} by $Z_\phi$.
The resulting denominator is  the renormalized inverse pair propagator in Eq.~\eqref{Pi:renorm}.
The product of the numerator factor $W_{\pi 0}(E,p)$ and $\sqrt{Z_\phi}$
 defines a renormalized numerator factor $W_\pi(E,p)$.
The resulting renormalized expression for  the $C=+$ S-wave transition amplitude is
\bea
\mathcal{A}_{s+}(E,p,p^\prime) &=& \mathcal{A}_\pi(E,p,p^\prime) 
+ W_\pi(E,p)\,  \frac{1}{ \mathcal{A}(E)^{-1} - \Pi(E)}\, W_\pi(E,p^\prime). 
\label{A-NLO:ren}
\eea
The renormalized numerator factor at NLO is
\beq
W_\pi(E,p) = 1 + g_\pi^2 \, G(E,p) + (R_0/\gamma)\, p^2/4  +\delta Z_\phi/2,
\label{Wpi:renorm}
\eeq
where $G(E,p)$ is given in Eq.~\eqref{G-E,p}.
We have used Eq.~\eqref{C2-r0} to set $C_2/C_0 = R_0/(2\gamma)$, where $R_0$  is a renormalized interaction parameter.
The renormalized self-energy $\Pi(E)$ at NLO  is given in Eq.~\eqref{PiNLO}, with $C_2/C_0$  replaced by $R_0/(2\gamma)$.

We proceed to verify that the terms in the renormalized  transition amplitude in Eq.~\eqref{A-NLO:ren}
are all UV finite at NLO.
We have already verified the cancellation of all the poles in $d-2$ and $d-3$
of the renormalized self-energy $\Pi(E)$ at NLO.
The function $G(E,p)$ has single poles in $d-2$ from the loop integrals $L_0$ and $J_1$.
It can be easily verified that the poles are cancelled by the $\delta Z_\phi/2$ term in Eq.~\eqref{Wpi:renorm},
where $\delta Z_\phi$ is the NLO term in the renormalization constant in Eq.~\eqref{Zphi:d->2}.
We conclude that the linear UV divergences in $W_\pi(E,p)$ cancel at NLO.

Having verified that the linear UV divergences in the numerator factors in Eq.~\eqref{A-NLO:ren}  cancel,
there is  nothing to be gained by making them explicit using power divergence subtraction.
We therefore choose to simplify the numerator factors by using conventional dimensional regularization
in which loop integrals are analytically continued to 
the neighborhood of $d=3$.  Since the function $G(E,p)$ has no poles in $d-3$,
we can simply set $d=3$ in the renormalized numerator factor $W_\pi$.

\subsection{Complex on-shell renormalization scheme}
\label{sec:RenormCOS}

In Ref.~\cite{Braaten:2015tga}, which introduced Galilean invariant XEFT, a renormalization prescription for 
$D^{*0} \bar D^0$ and $D^0 \bar D^{*0}$ amplitudes 
called the {\it complex on-shell} (COS) renormalization scheme was  introduced.
The $D^{*0} \bar D^0$ transition amplitude has a pole in the CM energy at the same complex energy $E_\mathrm{pole}$
as the complete pair propagator.
The pole energy is expressed in terms of the binding momentum $\gamma_X$  in Eq.~\eqref{Epole}.

The renormalization prescription for the COS scheme in  Ref.~\cite{Braaten:2015tga}
is that the pole in $E$ in the transition amplitude in Eq.~\eqref{A-NLO:ren} at $p = p^\prime = 0$
has the same value and the same  residue as at LO.
The pole energy $E_\mathrm{pole}$ has the same value if the LO binding momentum $\gamma$ is equal to  $\gamma_X$.
The condition for the pole at $p = p^\prime = 0$ to have the same residue is
\beq	
\frac{W^2_\pi(E,0)}{- \gamma_X + \kappa(E) - \Pi_\mathrm{COS}(E)}
\longrightarrow \frac{-\gamma_X/\mu}{E-E_\mathrm{pole}} 
\qquad \mathrm{as~} E \to E_\mathrm{pole},
\label{COSresidue}
\eeq
where $\Pi_\mathrm{COS}(E)$ is the renormalized pair self-energy in the COS scheme.
The two renormalization conditions for $\Pi_\mathrm{COS}(E)$ are
\begin{subequations}
\bea
\Pi_\mathrm{COS}(E_\mathrm{pole})& =& 0,
\label{COS1}
\\
\Pi_\mathrm{COS}^\prime(E_\mathrm{pole})  &=& \frac{\mu}{\gamma_X} \big[ W^2_\pi(E_\mathrm{pole},0) -1 )\big].
\label{COS2}
\eea
\label{COS1,2}%
\end{subequations}

We proceed to implement the COS scheme for the complete pair propagator at NLO.
The expansion to NLO of the numerator on the left side of Eq.~\eqref{COSresidue} is
\beq	
W^2_\pi(E,0) = 1  + 2 g_\pi^2 \, G(E,0) + \delta Z_\phi.
\label{Wpi^2NLO}
\eeq
The renormalized pair self-energy $\Pi_\mathrm{COS}(E)$ at NLO in the COS scheme  is given by 
Eq.~\eqref{PiNLO:renorm} with $C_2/C_0$ replaced by $R_0/(2\gamma)$
and with appropriate complex values for the counterterms $\delta C_0$ and $D_0$.
These counterterms correspond to subtractions proportional to $\kappa^0(E) = 1$ and $\kappa^2(E)$.
The solution to the renormalization conditions in  Eq.~\eqref{COS1,2} is
\bea
\Pi_\mathrm{COS}(E) &=& 
f_1 \, g_\pi^2\kappa_*   \left[\kappa(E) - \frac{\gamma_X}{2}  - \frac{1}{2\gamma_X}  \kappa^2(E) \right]
-    \delta Z_\phi \left[\kappa(E) - \gamma_X\right]
\nonumber\\
&&\hspace{-1.5cm} 
- \frac{g_\pi^2\,  G(E_\mathrm{pole},0)}{\gamma_X} \left[\kappa^2(E) - \gamma_X^2\right]
+\frac{R_0}{2 \gamma_X} \left[\kappa^3(E) +  \frac{\gamma_X^3}{2}  - \frac{3\gamma_X}{2}  \kappa^2(E) \right]
+g_\pi^2 \,  F_{4,\mathrm{sub}}(E),
\label{PiNLO:COS}
\eea
where $F_{4,\mathrm{sub}}(E)$ is obtained by subtracting from $F_4(E)$ in Eq.~\eqref{F4-E}
the first two  terms in its expansion in powers of $E-E_\mathrm{pole}$:
\beq
F_{4,\mathrm{sub}}(E)  = F_4(E) - F_4(E_\mathrm{pole})  - F_4^\prime(E_\mathrm{pole})  (E-E_\mathrm{pole}).
\label{F2sub-E}
\eeq
It is easy  to verify that the self-energy in Eq.~\eqref{PiNLO:COS}  vanishes at $E =E_\mathrm{pole}$
by using $\kappa(E_\mathrm{pole}) = \gamma_X$.
One can verify that its derivative with respect to $E$ at $E_\mathrm{pole}$ agrees with the NLO term 
on the right side of Eq.~\eqref{COS2} by also using $\kappa^\prime(E_\mathrm{pole}) = - \mu/\gamma_X$.
The additional  renormalization freedom associated with the renormalization constant $Z_\phi$ 
was not recognized in Ref.~\cite{Braaten:2015tga}, so $\delta Z_\phi$ was set to 0.

The expression for the renormalized self-energy $\Pi_\mathrm{COS}(E)$ in the COS scheme
in Eq.~\eqref{PiNLO:COS} is significantly more complicated than that for $\Pi(E)$ in the CT scheme in Eq.~\eqref{PiNLO:CT}.
The adjustable real interaction parameters in $\Pi_\mathrm{COS}(E)$
in Eq.~\eqref{PiNLO:COS} are $\mathrm{Re}[\gamma_X]$, $\mathrm{Im}[\gamma_X]$, and $R_0$.
There is no additional real interaction parameter 
analogous to $F_2$ in $\Pi(E)$ in Eq.~\eqref{PiNLO:CT}.
Such a parameter arises inevitably from the freedom in the choice of the finite real part 
accompanying the pole in $d-3$ of the counterterm $D_0$. 
The absence of such a term in Eq.~\eqref{PiNLO:COS} indicates that the
renormalization conditions in Eq.~\eqref{COS1,2} are insufficient.  
It is not clear how to extend these renormalization conditions to allow for the adjustable parameter $F_2$.
The parameters  $\mathrm{Re}[\gamma_X]$ and $\mathrm{Im}[\gamma_X]$ in $\Pi_\mathrm{COS}(E)$ are determined by
the real and imaginary parts of the pole energy $E_\mathrm{pole} =E_X -i \Gamma_X/2$.
The inputs $E_X$ and $\Gamma_X$ are not ideal  parameters,
because they are difficult to determine experimentally.
The only experimental determinations of the pole energy thus far are by the LHCb collaboration 
using the Flatt\'{e} model  \cite{Aaij:2020qga}, and their results do not have error bars.
One might as well use $\mathrm{Re}[\gamma]$ and $\mathrm{Im}[\gamma]$
 as the real adjustable parameters, as in the CT scheme.

\subsection{NLO scattering amplitude} 

We consider the elastic scattering of $D^{*0} \bar D^0$ and $D^{*0} \bar D^0$ in the CM frame
with  incoming relative momentum $\bm{p}$ and outgoing relative momentum $\bm{p}^\prime$.
Conservation of energy requires $p^\prime = p$.
The scattering angle $\theta$ is defined by  $\bm{\hat p} \cdot \bm{\hat p}^\prime = \cos \theta$.
The polarization vectors of the incoming and outgoing $D^{*0}$ 
are $\bm{\varepsilon}$ and $\bm{\varepsilon}'$.
The energy shell conditions require the total energy $E$ to have the complex value
\beq
E_p = E_* + p^2/(2 \mu).
\label{Ep}
\eeq
The T-matrix element for $D^{*0} \bar D^0 \to D^{*0} \bar D^0$ is obtained by multiplying 
the on-shell amputated connected Green function $ \mathcal{T}^{ij}(E_p,\bm{p},\bm{p}^\prime)$
by the external line factors in Eq.~\eqref{D*external} for the incoming and outgoing $D^{*0}$:
\beq
{\cal T}(\bm{p}, \bm{\varepsilon};\bm{p}^\prime,\bm{\varepsilon}^\prime) =  
\sum_{ij} \mathcal{T}^{ij}(E_p,\bm{p},\bm{p}^\prime) 
 \varepsilon^i  {\varepsilon'}{}^j{}^* .
\label{T-NLO}
\eeq
The T-matrix element for scattering in the $C=+$ channel is the sum of the T-matrix elements for
$D^{*0} \bar D^0 \to D^{*0} \bar D^0$, $D^{*0} \bar D^0 \to D^0 \bar D^{*0}$,
and $D^0 \bar D^{*0} \to D^{*0} \bar D^0$, $D^0 \bar D^{*0} \to D^0 \bar D^{*0}$ multiplied by 1/2.
It can be projected onto the S-wave channel by averaging over the directions of the momenta $\bm{p}$ and  $\bm{p}^\prime$.

The T-matrix element for S-wave scattering in the $C=+$ channel 
can be expressed in terms of the scalar  transition amplitude $\mathcal{A}_{s+}$ 
in Eq.~\eqref{A-NLO:ren} evaluated on shell by setting $p^\prime = p$ and $E=E_p$:
\beq
\mathcal{T}_{s+}(p) = (2\pi/\mu)\, \mathcal{A}_{s+}(E_p,p,p) \, .
\label{T-Swave}
\eeq
The T-matrix element for S-wave scattering in the $C=+$ channel at NLO is
\beq
\mathcal{T}_{s+}(p) = (2\pi/\mu)
\left( \mathcal{A}_\pi(E_p,p,p) +  \frac{W_\pi^2(E_p,p)}{(-\gamma - i\, p) - \Pi(E_p)} \right) .
\label{T-SwaveNLO}
\eeq
The  tree-level pion-exchange term is obtained by evaluating Eq.~\eqref{Api-S} on shell:
\beq
\mathcal{A}_\pi(E_p,p,p)  = \frac{r g_\pi^2}{6\pi  \sqrt{1-r}} 
 \Bigg( \frac{r \big[(2-r)\kappa_*^2 - r \, p^2 \big]}{4 \sqrt{1-r} \, p^2}
 \log \frac{r \kappa_*^2 - \big(1+\sqrt{1-r}\, \big)^2 p^2}{r \kappa_*^2  - \big(1- \sqrt{1-r}\,\big)^2 p^2}
+ (2-r) \Bigg).
\label{ApiS-onshell}
\eeq
The  numerator factor is  obtained by evaluating $G(E,p)$ in Eq.~\eqref{G-E,p} on shell 
and inserting it into Eq.~\eqref{Wpi:renorm}.
The function $G(E,p)$ reduces on shell to
\bea
G(E_p,p)  &=& 
 i\, \frac{r }{6\pi  \sqrt{1-r}} 
 \Bigg( \frac{r \big[(2-r)\kappa_*^2 -rp^2 \big]}{2 \sqrt{1-r} \, p}
 \log \frac{\sqrt{r}\,  \kappa_* + \big(1+ \sqrt{1-r}\, \big) p}{\sqrt{r}\,  \kappa_* + \big(1- \sqrt{1-r}\, \big) p}
\nonumber\\
&& \hspace{3cm}
+ (2-r) p - r^{3/2} \kappa_* \bigg).
\label{G-onshell}
\eea
The renormalized pair propagator at $E = E_p$
can be obtained by  replacing  $\kappa(E)$ by $- i p$ in Eq.~\eqref{PiNLO:CT}:
\bea
\Pi(E_p)  &=&   \big( \delta Z_\phi -f_1g_\pi^2 \kappa_* \big)\, i\,p-  \big( F_2+ i\, \mathrm{Im}[f_2] \big) g_\pi^2 \, p^2
+ i \frac{R_0}{2\gamma} p^3 + g_\pi^2 \,  F_4(E_p).
\label{Pi-onshell}
\eea
In the CT scheme, the coefficient of  $i\, p$ is 0.

\subsection{Breakdown of the effective range expansion} 

In the case of only short-range interactions,
a scattering amplitude can be expanded in powers of  the relative momentum.
This expansion is called the {\it effective range expansion}.
The scattering length $a$ and the effective range $r_e$ can be
defined as coefficients in the expansion of the reciprocal of the T-matrix  element $\mathcal{T}(p)$
in powers of the momentum $p$: 
\beq
\frac{2\pi/\mu}{\mathcal{T}(p)} = 
- \frac{1}{a} - i \, p +\frac12 r_e\,   p^2 + \mathcal{O}(p^4).
\label{T-NLO:as+}
\eeq
Unitarity requires that the only odd power of $p$ in the expansion is  the $- ip$ term.

At LO, the $C=+$ S-wave transition amplitude reduces to  the  pair propagator $\mathcal{A}(E)$ in Eq.~\eqref{A0-E:3}.
The T-matrix element for S-wave scattering in the $C=+$ channel at LO is therefore
\beq
{\cal T}_{s+,\mathrm{LO}}(p) = \frac{2 \pi/\mu}{-\gamma- i \,p} .
\label{T-LO}
\eeq
Comparing with Eq.~\eqref{T-NLO:as+},
we see the inverse scattering length $1/a_{s+}$ in the $C=+$ S-wave channel at LO is equal to $\gamma$  
and the effective range at LO is zero.

We proceed to consider the expansion of  the reciprocal of the $C=+$  S-wave T-matrix  at NLO in powers of $p$.
The  tree-level pion-exchange term in Eq.~\eqref{ApiS-onshell} has an expansion in powers of $p^2$.
It reduces at small $p$  to
\beq
\mathcal{A}_\pi(E_p,p,p) = - \frac{2\sqrt{1-r}\,  g_\pi^2}{3\pi \kappa_*^2}\, p^2+ \mathcal{O}(p^4).
\label{ApiS-onshell:p0}
\eeq
The  numerator factor in Eq.~\eqref{T-SwaveNLO} has an expansion in powers of $p$.
It reduces  at small $p$  to
\bea
W_\pi(E_p,p)  &=& 1 + \frac12\left( \delta Z_\phi + i\, \frac{2\sqrt{1-r}\,  r^{3/2}}{3 \pi} g_\pi^2 \kappa_* \right)
+ \left( \frac{R_0}{4 \gamma} + i\, \frac{(4+r)  \sqrt{1-r}\, r^{1/2}}{9\pi} \, \frac{g_\pi^2}{\kappa_*} \right) p^2
\nonumber\\
&&\hspace{0.5cm} 
-i\, \frac{2\sqrt{1-r}\,  g_\pi^2}{3\pi \kappa_*^2}\, p^3 + \mathcal{O}(p^4).
\label{N-onshell:p0} 
\eea
Note that  there is no term linear in $p$.
The inverse scattering length $1/a_{s+}$ can be obtained by taking the $p \to 0$ limit of 
$\mathcal{T}_{s+}$ in Eq.~\eqref{T-SwaveNLO}.
The expansion of $1/a_{s+}$  to NLO in $g_\pi^2$  is
\beq
1/a_{s+} = \left(1- \delta Z_\phi  - i\,  \frac{2 \sqrt{1-r}\, r^{3/2}}{3 \pi} g_\pi^2 \kappa_*\right) \gamma .
\label{as+NLO}
\eeq
In the CT scheme, $\delta Z_\phi$ is replaced by the correction to 1 in Eq.~\eqref{ZphiCT}.

In the expansion of $(2\pi/\mu)/\mathcal{T}_{s+}(p)$ in powers of $p$,
the term linear in $p$ differs  from the term $-i\, p$ in Eq.~\eqref{T-NLO:as+}
that is required by unitarity if all interactions have short range.
This breakdown of the effective range expansion can be attributed to the effects of the successive exchange of pions 
that are almost on their energy shell.
The coefficient of $-i\,p$ differs from 1 by a term that is almost pure imaginary.
The coefficient can therefore be expressed as the NLO approximation to a factor that is almost a complex phase.
This complex phase can be interpreted as the phase shift from the successive exchange of pions.
If the complex phase factor is  factored out of the expression for $(2\pi/\mu)/\mathcal{T}_{s+}(p)$,
the remaining factor
has an expansion in powers of $p$ 
with the linear term $-i\, p$ as in  Eq.~\eqref{T-NLO:as+}:
\beq
\frac{2\pi/\mu}{\mathcal{T}_{s+}(p)} \approx 
\exp\left(-i\, \frac{(2-r) r^{3/2}}{3 \pi \sqrt{1-r}} g_\pi^2 \kappa_* \right)
\left[- \gamma_{s+} - i\, p + \frac12 r_{s+}\, p^2 + \mathcal{O}(p^3) \right].
\label{1/T-NLO}
\eeq
In the CT scheme, the coefficient $\gamma_{s+}$ is just $\gamma$.
The coefficient $r_{s+}$ is
\bea
r_{s+} &=& R_0 + 2 F_2 \, g_\pi^2 + \frac{4}{3 \pi} \bigg[ -i\, \frac{\big(1+\sqrt{1-r}\,  \big) r^{5/2}}{4\sqrt{1-r}}
+ i\, \frac{r^3}{4(1-r)} \arccos\big( \sqrt{r}\,  \big)
\nonumber\\
&&\hspace{4cm}
+  i \frac{(4+r)\sqrt{1-r}\,  r^{1/2}}{3}  \frac{\gamma}{\kappa_*}+ \sqrt{1-r}\left( \frac{\gamma}{\kappa_*} \right)^2 \bigg]  g_\pi^2.
\label{rs+} 
\eea
The contributions to $r_{s+}$ that are almost pure imaginary
could alternatively be absorbed into terms proportional to $p^2$ in the  phase shift.
At NLO, $1/\mathcal{T}_{s+}(p)$ has a well-behaved effective range expansion though order $p^2$ 
modulo an overall phase factor.
It  would be interesting to know if this remains true at higher orders.

In Ref.~\cite{Jansen:2013cba}, Jansen {\it et al.}\  pointed out that the effective range expansion
for $D^{*0} \bar D^0$ scattering breaks down beyond LO in XEFT,  
because of the effects of the exchange of a pion that can be on its energy shell.
They argued that the S-wave $D^{*0} \bar D^0$ scattering length 
remains well defined, but that the breakdown of the effective range expansion
made the effective range undefined.
 Jansen {\it et al.}\   calculated the scattering length to NLO in original XEFT,
truncating the expression at first order in an expansion 
in powers of $\gamma_X/\sqrt{2m \delta}$ and at leading order in  $m/M$
or, equivalently, $r \approx 2m/M$ \cite{Jansen:2013cba}.
Their result for $1/a_{s+}$ depends on a renormalization scale $\Lambda$ through
terms of the form $1/(\gamma- \Lambda)$.
These terms were produced by an infrared resummation that was apparently incompatible with their renomalization prescription.

In Ref.~\cite{Braaten:2015tga}, the inverse scattering length at NLO was calculated 
using Galilean-invariant XEFT  in the COS renormalization scheme.
Using the result for $\Pi_\mathrm{COS}(E)$ in Eq.~\eqref{PiNLO:COS} with $\kappa(E)$ set to 0
and the result for $W_\pi(E_*,0)$ from Eq.~\eqref {Wpi^2NLO}, 
the  inverse scattering length at NLO can be expressed as
\bea
\frac{1}{a_{s+}} &=& 
\gamma_X
+  g_\pi^2 \left( G(E_\mathrm{pole},0) - 2 G(E_*,0) -\frac{f_1}{2} \kappa_* \right)\gamma_X
\nonumber \\
&& \hspace{1cm}
+\frac14 R_0\gamma_X^2
-g_\pi^2  \left( F_4(E_\mathrm{pole}) 
+ F_4^\prime(E_\mathrm{pole})  \frac{\gamma_X^2}{2 \mu} \right),
\label{as+}
\eea
where $F_4(E)$ is given in Eq.~\eqref{F4-E} and $G(E,0)$ is 
\bea
G(E,0) &=&
i\,  \frac{\sqrt{1-r}\, r}{3\pi}  
 \frac{ (2r\mu E)^{3/2} - [2\mu(E-E_*) ]^{3/2} }{ (2r\mu E) - [2\mu(E-E_*)]}.
 \label{G-E,0}
\eea
Note that $1/a_{s+}$ in the COS scheme
does not depend on the choice for the renormalization constant $Z_\phi$  for the pair propagator.
This result in Eq.~\eqref{as+} is much more complicated than that in the CT scheme in Eq.~\eqref{as+NLO}.
The inverse scattering length cannot be calculated analytically because of the $F_4$ terms
in Eq.~\eqref{as+}.
Those terms have expansions in powers of $\gamma_X$ that begin at order $\gamma_X^4$.
An analytic expression for $1/a_{s+}$ in the COS scheme can be obtained 
as an expansion in powers of $\gamma_X$.
The expansion through fourth order in $\gamma_X$ is
\bea
\frac{1}{a_{s+}} &=&\left( 1  -i \, \frac{(2-r) \, r^{3/2}}{6\pi \sqrt{1-r}} g_\pi^2 \kappa_* \right)  \gamma_X
+\frac14 R_0 \gamma_X^2
\nonumber\\
&&
- i\, \frac{(2+r)\sqrt{1-r}\, r^{1/2}}{6\pi } \frac{g_\pi^2\gamma_X^3 }{\kappa_*}
+ \left(  f_4 -  \frac{\sqrt{1-r}}{3\pi }  \right) \frac{g_\pi^2\gamma_X^4 }{\kappa_*^2} ,
\label{as+:expand}
\eea
where $f_4$ is given in Eq.~\eqref{f4-anal}.
The coefficient of each power of $\gamma_X$ in Eq.~\eqref{as+:expand}
can be expanded in powers of the small parameter $r$ of XEFT.
The expansion of $1/a_{s+}$  truncated after the $\gamma_X^4$ term reduces to 
\bea
1/a_{s+} &=& \gamma_X
+ \frac14 R_0 \gamma_X^2
+ \frac{1}{3 \pi} g_\pi^2\kappa_*^2
\Bigg[ 
- i \, r^{3/2} \left(1 +  \frac18 r^2 \right)  \frac{\gamma_X}{\kappa_*}
\nonumber \\
&&\hspace{1cm}
- i\,  r^{1/2}\left(1 - \frac38 r^2 \right) \left(\frac{\gamma_X}{ \kappa_*}\right)^3
+ \left(1 - \frac12 r - \frac18 r^2 \right)  \left(\frac{\gamma_X}{\kappa_*}\right)^4 \Bigg] .
\label{as+:gammaX}
\eea
The coefficient of each term has been expanded to relative order  $r^2$.
In each of the three NLO correction terms proportional to $g_\pi^2$  that are shown,  
the sum of the power of $\gamma_X$ and the leading power of $\sqrt{r}$ is equal to 4.
The higher powers of $\gamma_X/\kappa_*$
are therefore partly compensated for by the fewer powers of $\sqrt{r}$.
In  Ref.~\cite{Braaten:2015tga}, there is an error in the result for $1/a_{s+}$ in the COS scheme:
the coefficient of $\gamma_X/\kappa_*$
has a factor $1- \tfrac12 r +\tfrac18 r^2$ instead of $1 +\tfrac18 r^2$. 

If we take the limit $g_\pi^2 \to 0$, the expansion of the reciprocal of the NLO T-matrix element  in the CT scheme
at small $p$ reduces to
\beq
\frac{2\pi/\mu}{\mathcal{T}_{s+}(p)} = 
- \gamma - i\, p + \frac{R_0}{2} p^2 - \frac{3R_0^2}{16 \gamma} p^4 + i\,  \frac{R_0^2}{16 \gamma^2} p^5
 + \mathcal{O}(p^6).
\label{1/T-NLO:nopi}
\eeq
The $p^5$ term indicates an obvious breakdown of the effective range expansion, but
the negative power of $\gamma$ in the coefficient of the $p^4$ term is another indication.
Comparison with the  effective range expansion in Eq.~\eqref{T-NLO:as+}
reveals that $R_0$ has a simple interpretation in the CT scheme. 
It is the effective range in the limit in which pion interactions are turned off.

\subsection{Pion-exchange resummation} 

We have found that pion exchange causes a breakdown of the effective range expansion
for the T-matrix element for $D^{*0} \bar D^0$ scattering in the S-wave $C=+$ channel.
In the case of strong  short-range interactions plus weak long-range interactions,
the effective range expansion can be modified in various ways.
The simplest possible modification is additional odd powers of $p$ beginning at order $p^3$.
Some of the odd powers of $p$ could be factored out into an overall phase shift.
However the modifications could be much more dramatic.
An extreme case of a long-range interaction is the Coulomb interaction between charged particles.
In this case, it is necessary to resum the effects of Coulomb interaction to all orders.
The resummation of Coulomb interactions in low-energy proton-proton scattering was first treated 
in an effective field theory framework by Kong and Ravndal {\it et al.}\ \cite{Kong:1999sf}.
The formalism was extended in Ref.~\cite{Braaten:2017kci} to a two-channel system of dark matter particles
in which one channel is a pair of charged particles and the other channel is a pair of neutral particles.
The T-matrix element in the S-wave channel for a single pair of charged particles with strong short-range interactions 
has the form
\begin{equation}
 \mathcal{T}(p)= (2\pi/\mu)\, 
\big[  \mathcal{A}_C(p) + W_C^2(p)\,  \mathcal{A}_s(p) \big] .
\label{eq:A-sC}
\end{equation}
For a pair of charged particles with opposite unit electric charges,
the resummation of Coulomb interactions without any short-range interactions
gives the S-wave Coulomb  amplitude $\mathcal{A}_C(p)$:
\begin{equation}
\label{eq:ACoulomb}
\mathcal{A}_C(p) = 
\left( 1 - \frac{\Gamma(1-i \alpha \mu/p)}{\Gamma(1+i \alpha \mu/p)}  \right) \frac{i\, \pi}{\mu \, p} ,
\end{equation}
where $\alpha$ is the fine-structure constant of QED.
The resummation of Coulomb interactions before the first short-range interaction
or after the last  short-range interaction gives the amplitude $W_C(p)$, whose square is:
\begin{equation}
W_C^2(p) =  \frac{2 \pi \alpha \mu/p}{1 - \exp(-2 \pi \alpha \mu/p)} \, 
\frac{\Gamma(1-i \alpha \mu/p)}{\Gamma(1+i \alpha \mu/p)}   .
\label{eq:S-eta}
\end{equation}
The amplitude  $\mathcal{A}_s(p)$ in Eq.~\eqref{eq:A-sC} comes from short-range interactions only.
The T-matrix element from this term only would presumably have a conventional effective range expansion
analogous to that in Eq.~\eqref{1/T-NLO}.

The T-matrix element for $D^{*0} \bar D^0$ scattering in the $C=+$ S-wave channel in Eq.~\eqref{T-SwaveNLO}
has the same form as that for strong short-range interactions plus Coulomb interactions in Eq.~\eqref{eq:A-sC}.
The analogous off-shell $C=+$ S-wave transition amplitude is given in Eq.~\eqref{A-NLO:ren}.
Each of the three terms  in Eq.~\eqref{A-NLO:ren}  has been calculated to NLO in the XEFT power-counting.
The only diagram that contributes to $\mathcal{A}_\pi(E,p,p^\prime)$ at NLO 
is one in which a pion is exchanged between the charm mesons.
The NLO correction to $W_\pi(E,p)$ comes from a diagram with the exchange of a  pion.
The accuracy of the T-matrix element could be improved by calculating all three terms  in Eq.~\eqref{T-SwaveNLO} to NNLO.
The NNLO contributions to $\mathcal{A}_\pi$ and $W_\pi$ 
come from diagrams with two successive pion exchanges.
Since the pions that are exchanged can be on shell,
the terms $\mathcal{A}_\pi$ and $W_\pi$ include effects from much longer distances
than the pair self-energy $\Pi(E)$.
It is possible that an accurate calculation of  the T-matrix element 
would require the resummation of successive pion exchanges to all orders in $\mathcal{A}_\pi$ and in $W_\pi$.

\section{Outlook}
\label{sec:Outlook}

As an effective field  theory for a sector of QCD that includes the $X(3872)$,
XEFT allows systematically improvable calculations of some of the properties of this resonance.
In the original formulation of XEFT, the interactions of the charm mesons with pions were chosen to
have a form motivated by the approximate chiral symmetry of QCD  \cite{Fleming:2007rp}.
The Galilean-invariant  formulation of XEFT developed in Ref.~\cite{Braaten:2015tga} was a significant improvement, 
because the Galilean symmetry constraints the ultraviolet divergences and it significantly simplifies analytic results.
We have introduced a new formulation of Galilean-invariant XEFT with a dynamical pair field
that annihilates a pair of charm mesons in the resonant channel.
The new formulation simplifies calculations at next-to-leading order by making some cancellations 
of UV divergences between diagrams automatic. 
The terms in the Lagrangian for this formulation of XEFT are given in Section~\ref{sec:Lagrangian}
and the Feynman rules are given in Section~\ref{sec:Frules}.
We also introduced a new renormalization scheme called the complex threshold (CT) scheme
that makes analytic results at next-to-leading order much simpler than with the 
complex on-shell (COS) scheme introduced in  Ref.~\cite{Braaten:2015tga}.
The advantages of the CT scheme were illustrated with NLO calculations of the 
complex pole energy of $X$ in Section~\ref{sec:PairPropNLO}
and the elastic $D^{*0} \bar D^0$ scattering amplitude in Section~\ref{sec:D*Dscat}.

An important insight  provided by our new formulation of Galilean-invariant XEFT
is that there is an additional interaction parameter at NLO that was not recognized in Ref.~\cite{Braaten:2015tga}.
In the threshold expansion of the pair self-energy in powers of $\kappa(E)$,
the CT scheme requires the total subtraction of the terms proportional to $\kappa^0(E)$ and $\kappa(E)$.
Renormalization at NLO also requires a partial subtraction of the term proportional to $\kappa^2(E)$.
The freedom in the choice of the  finite part of that subtraction leads to the real interaction parameter $F_2$
in the renormalized pair self-energy in Eq.~\eqref{PiNLO:CT}.
The other adjustable real interaction parameters at NLO are  the real and imaginary parts of $\gamma$,
the effective range $R_0$ in the absence of pion interactions, and the strength of the 
$D^{*0}\bar D^0$ to $D^0\bar D^0 \pi^0$ transition.

Another insight  provided by our new formulation of Galilean-invariant XEFT is that renormalization of the 
$D^{*0}\bar D^0$ scattering amplitude requires the pair renormalization constant $Z_\phi$.
At NLO, the UV divergences  canceled by this renormalization constant are linear UV divergences.
The need for this renormalization was not recognized in Ref.~\cite{Braaten:2015tga},
because conventional dimensional regularization sets linear ultraviolet divergences to 0.
If power divergence subtraction had been used to make the  linear UV divergences explicit,
the dependence on the renormalization scale of both the numerator and the denominator 
of the resonant term in the $D^{*0}\bar D^0$ scattering amplitude in Eq.~\eqref{T-SwaveNLO}
would have made the failure of the renomalization procedure evident.

Numerical calculations of the momentum distribution for the decay of $X$ into $D^0 \bar D^0 \pi^0$ 
using original XEFT at NLO  have revealed that the NLO corrections are surprisingly large 
\cite{Fleming:2007rp,Dai:2019hrf}.
The power counting rules of XEFT guarantee that calculations can be systematically  improved, 
but the large NLO corrections raise the issue of whether  
the systematic expansions converge fast enough to provide useful quantitative approximations.
Our new complex threshold (CT)  renormalization scheme for  $D^*\bar D$ and  $D\bar D^*$
amplitudes provides a possible solution to the problem of large NLO corrections.
The NLO corrections to the decay rate for $X$ into $D^0 \bar D^0 \pi^0$
from pion emission were calculated numerically in Ref.~\cite{Fleming:2007rp} as functions of the LO binding momentum,
and they are surprisingly large even for tiny values of $\gamma$.
However pion emission also gives imaginary corrections to the binding momentum of $X$.
Some of the large NLO corrections can be attributed to expanding LO results to first order 
in the corrections to the binding momentum.
In the CT scheme, these corrections are absorbed into the parameter $\gamma$ itself.
The calculation of the momentum distribution for the decay of $X$ into $D^0 \bar D^0 \pi^0$
at NLO using the CT scheme would involve a subtraction of part of the NLO corrections
in Refs.~\cite{Fleming:2007rp,Dai:2019hrf}.  The resulting NLO corrections are likely to be much smaller.
It would be very useful to have an analytic calculation of the decay rate of $X$ into $D^0 \bar D^0 \pi^0$ 
at NLO using the CT scheme.
If the NLO corrections are much smaller than those in Refs.~\cite{Fleming:2007rp,Dai:2019hrf},
it would provide convincing evidence that XEFT
gives approximations that are not only  systematically improvable but also quantitatively useful.

Jansen {\it et al.}\   pointed out 
that  the effective range expansion for $D^{*0} \bar D^0$ scattering in XEFT breaks down at NLO 
from the effects of the exchange of pions that can be on shell \cite{Jansen:2013cba}.
Our calculation of the T-matrix element for $D^{*0} \bar D^0$ scattering in the $C=+$ S-wave channel 
to NLO makes the breakdown of the effective range expansion explicit.
Its form differs  from that required by unitarity for a system with short-range interactions only already at order $p$.
The breakdown of the effective range expansion raises the issue of whether the power counting rules of XEFT 
provide a systematically improvable approximation for this scattering amplitude.
It could be that an accurate approximation requires resumming the effects of successive pion exchanges
to all orders in both the amplitude $\mathcal{A}_\pi$ for scattering through pion exchange only
and  in the amplitude $W_\pi$ that takes into account  the effects of pion exchange after the last pair amplitude.
The analogous amplitudes in the case where the long-range interactions are successive Coulomb interactions 
have been calculated analytically.
The analytic calculation of these  amplitudes in the case where the long-range interactions 
come from successive exchanges of pions that can be on shell is a challenging problem.

XEFT was originally presented as an effective field theory for the sector of QCD consisting of 
$D^*\bar D$, $D \bar D^*$, and $D \bar D \pi$, and $X$ with total energy  
near the $D^*\bar D$ threshold \cite{Fleming:2007rp}.
As pointed out in  Ref.~\cite{Braaten:2010mg},
XEFT can also be applied to the sector of QCD consisting of 
$D^*\bar D^*$, $D^*\bar D \pi$, $D \bar D^*\pi$, $D \bar D \pi \pi$, and $X \pi$
with total energy  near the $D^*\bar D^*$ threshold. 
It was used to calculate cross sections for  $D^* \bar D^* \to \pi X$ and $\pi X \to D^* \bar D^*$  \cite{Braaten:2010mg}.
XEFT has also been used to calculate cross sections for $DX$ scattering and $D^*X$ scattering \cite{Canham:2009zq}.
One interesting application of XEFT in the sector of QCD that includes $X\pi$ is to 
the calculation of charm-meson triangle singularities
in the production of $X \pi$ or $X \gamma$ from $D^*\bar D^*$ created at short distances.
Back in 2006, Dubynskiy and Voloshin pointed out the existence of a peak in the cross section for
$e^+ e^-$ annihilation into $X \gamma$ near the $D^{*0}\bar D^{*0}$ threshold \cite{Dubynskiy:2006cj}.
Braaten, He, and Ingles pointed out  the existence of peaks in the production rate of $X\pi$ 
near the $D^*\bar D^*$ threshold in $B$ meson decay into $K X \pi$ \cite{Braaten:2019yua}
and in the prompt production of $X \pi$ at hadron colliders \cite{Braaten:2019sxh}.
They did not immediately recognize that the peaks were produced by triangle singularities.
Guo was the first to point out that charm-meson triangle singularities can produce peaks in  
$X \gamma$ production rates near the $D^{*0}\bar D^{*0}$ threshold \cite{Guo:2019qcn}.
He suggested that their line shapes could be used for a precise determination of the mass of $X$.
Braaten, He, and Ingles presented a thorough study of $e^+ e^-$ annihilation into $X \gamma$ 
near the $D^{*0}\bar D^{*0}$ threshold that was informed by the triangle singularity \cite{Braaten:2019gfj}.
The effects of the charm-meson triangle singularity on the momentum distributions for $D^{*0} \bar{D}^0 + \gamma$
were studied in Ref.~\cite{Braaten:2020iye}.
There have been several other recent studies of the effects of triangle singularities 
on the production of $X$ \cite{Sakai:2020ucu,Molina:2020kyu,Sakai:2020crh}.
All of the studies thus far have been carried out using methods equivalent to XEFT at LO.
The methods developed in this paper could be used to extend the accuracy to NLO.

Although it is not widely recognized in the high energy physics community,
the applicability of XEFT is not limited to  a charm-meson molecule whose 
energy has been tuned to near the $D^{*0}\bar D^0$ threshold.
It applies equally well to the $\chi_{c1}(2P)$ charmonium state or to a compact tetraquark whose 
energy has been tuned to near the $D^{*0}\bar D^0$ threshold.
In any of these cases, the resonant interactions with the pairs of charm mesons $D^{*0}\bar D^0$ 
and $D^0 \bar D^{*0}$ transform the meson into a large weakly-bound molecule with the 
particle content in Eq.~\eqref{X-DD}.
The differences between the various possibilities for the state when its energy is not tuned to 
near the $D^{*0}\bar D^0$ threshold can only be taken into account  through the parameters
in the effective Lagrangian for XEFT.
The development of a formulation of XEFT that is convenient for calculations beyond leading order
provides  motivation for deducing constraints on the parameters of XEFT that  reflect 
the various possibilities for its nature in the absence of the fine tuning of its energy.

\begin{acknowledgments}
This work was supported in part by the U.S.\ Department of Energy under grant DE-SC0011726,
the National Natural Science Foundation of China under grant 11905112, 
the Natural Science Foundation of Shandong Province under grant ZR2019QA012, 
and the Fundamental Research Funds of Shandong University under grant 2019GN038.
\end{acknowledgments}

\begin{appendix}

\section{Loop Integrals}
\label{sec:LoopInt}

The loop integrals required to calculate the NLO transition amplitude using dimensional regularization
were evaluated in Appendix~A of Ref.~\cite{Braaten:2015tga}.
There were errors in some of the loop integrals.
In this Appendix, we present  all the relevant integrals,
and we point out some of the errors in Ref.~\cite{Braaten:2015tga}. 

The loop integrals are functions of an energy $E$ that is near the complex threshold energy $E_*$
defined in Eq.~\eqref{E*}.
The integrals have  branch cuts that must be taken into account to evaluate the integrals correctly.
The energy $E_*$, which has a small negative imaginary part,
is actually on the second sheet of a complex variable  with a branch point at 0.
This can be emphasized by writing $E_*$ in the form
\beq	
E_* = e^{-2\pi i}\,  \big(\delta -i \Gamma_{*0}/2 \big),
\label{E*-branch}
\eeq
where the second factor is on the first sheet.
Thus $E_*$ is close to the energy $\delta+i \epsilon$ in the limit $\epsilon\to 0^+$, which is on the first sheet.
If $- 2 \mu E_*$ is raised to a variable power $p$, the correct branch is given by
\beq	
\big (- 2 \mu E_*\big)^p = e^{-i\, p\, \pi}\,  \kappa_*^{2p},
\label{-2muE*-branch}
\eeq
where $\kappa_*$ is the square root of a variable on the first sheet:
\beq	
\kappa_* = \sqrt{2\mu(\delta -i \Gamma_{*0}/2)}.
\label{kappa*}
\eeq

\subsection{One-loop momentum integrals}
\label{sec:1loop}

The 1-loop momentum integrals whose integrands are a $D^*$ propagator raised to an integer power are
\beq
J_n(E) = \int_{\bm{p}} \frac{1}{[ p^2 - 2\mu (E - E_*)]^n} .
\label{int-Jn}
\eeq
The analytic result for this integral in $d$ dimensions is
\beq
J_n(E)  =\frac{\Gamma(n-d/2)}{(4 \pi)^{d/2}\Gamma(n)} 
 \Lambda^{3-d} \, \big[ 2 \mu (E_* -E) \big]^{d/2-n}.
\label{Jn}
\eeq
In Ref.~\cite{Braaten:2015tga}, 
the factor of $\Gamma[n]$ in the denominator was omitted.  
The 1-loop momentum integrals whose integrands are a $\pi$ propagator raised to an integer power are
\beq
I_n(E)=\int_{\bm{p}} \frac{1}{[p^2 - 2\mu_\pi E - i \epsilon]^n},
\label{int-In}
\eeq
where $\mu_\pi$ is the $\pi D$ reduced mass.
The analytic result for this integral in $d$ dimensions is
\beq
I_n(E)  = 
 \frac{ \Gamma(n-d/2)}{(4 \pi)^{d/2}\Gamma(n)} \Lambda^{3-d} 
\left[ -  2 \mu_\pi E \right]^{d/2-n}.
\label{In}
\eeq
In Ref.~\cite{Braaten:2015tga},  
the factor of $\Gamma[n]$ in the denominator was omitted.  
The integral in Eq.~\eqref{int-In} defines an analytic function of $E$ with a branch cut along the positive real axis.
If the real part of $E$ is positive, the branch of the function can be selected by
replacing the minus sign inside the square brackets by the appropriate complex phase.

The 1-loop momentum integrals whose integrands have a $D^*$ propagator raised to an integer power
and a single $\pi$ propagator are
\beq
L_n(E,p) = \int_{\bm{q}} 
\frac{1}{[q^2 - 2\mu (E - E_*)]^n}  \, 
\frac{1/(2 \mu_\pi)}{ (\bm{p} + \bm{q})^2/(2m) + (p^2 + q^2)/(2M) - E -  i \epsilon} .
\label{int-Ln}
\eeq
The function $L_0$ can be expressed in terms of the integral $I_1$ given by Eq.~\eqref{In}:
\beq
L_0(E,p) =  I_1(E- p^2/(2 \mu)).
\label{L0}
\eeq
The function $L_1$ can be expressed as a Feynman parameter integral:
\bea
L_1(E,p) &=& 
\Lambda^{3-d}\frac{\Gamma(2-d/2)}{(4 \pi)^{d/2}} 
\int_0^1 {\rm d}x\,   \Big( 2x\mu E_* + ( x+(1-x)r )  [ -2 \mu E + (1-x)  p^2] \Big)^{d/2 -2} ,
\nonumber
\\
\label{L1}
\eea
where $r = \mu_\pi/\mu$ is the reduced-mass ratio  in Eq.~\eqref{r-mu}.
In Ref.~\cite{Braaten:2015tga},  there were errors in the expression for this integral.

\subsection{Two-loop momentum integrals}
\label{sec:2loop}

The 2-loop momentum integrals whose integrands have one or two $D^*$ propagators 
and a  $\pi$ propagator all raised to integer powers are
\bea
K_{lmn}(E) &=& \int_{\bm{p}} \int_{\bm{q}} 
\frac{1}{[ p^2 - 2\mu (E - E_*)]^m [q^2 - 2\mu (E - E_*)]^n} 
\nonumber
\\
&& \hspace{1cm}
\times  \frac{(2\mu)^{-l}}{[ (\bm{p} + \bm{q})^2/(2m) + (p^2 + q^2)/(2M) - E - i \epsilon]^l} .
\label{int-Klmn}
\eea
The specific integrals that appear in the NLO pair propagator are $K_{110}$, $K_{120}$, and  $K_{111}$.
The  integrals with a single $D^*$ propagator can be expressed as integrals over a single Feynman parameter:
\begin{subequations}
\bea
K_{110}(E)  &=&
\frac{ \Gamma(2-d)\, r^{d/2}}{(4 \pi)^d}  \Lambda^{6-2d}
\int_0^1 {\rm d}x\,  (1-x)^{-d/2}
[ 2 \mu (x E_* - E) ]^{d -2},
\label{K110-E:d}
\\
K_{120}(E)  &=&
\frac{ \Gamma(3-d)\, r^{d/2}}{(4 \pi)^d}  \Lambda^{6-2d}
\int_0^1 {\rm d}x\,  x (1-x)^{-d/2}
[ 2 \mu (x E_* - E) ]^{d-3}.
\label{K120-E:d}
\eea
\label{K110,K120-E:d}%
\end{subequations}
The integrals with two $D^*$ propagators can be expressed as an integral over two Feynman parameters:
\bea
K_{111}(E)  &=&
\frac{\Gamma(3-d)\, r^{d/2} }{(4 \pi)^d} \Lambda^{6-2d} 
\int_0^1 {\rm d}w\,  w\,  \big[ 2 \mu (w E_* - E) \big]^{d-3}
\nonumber
\\
&& \hspace{4cm} 
\times \int_0^1 {\rm d}t\,  
\left[ 1-w + r w^2 t(1-t)\right]^{-d/2}.
\label{K111-E:d}
\eea

\subsection{Poles in $\bm{d-2}$}
\label{sec:poles2}

In a dimensionally regularized loop integral, poles in $d-2$ are associated with 
linear ultraviolet (UV) divergences in 3 spatial dimensions.  
The poles in $d-2$ of the loop integrals were determined in Appendix A of Ref.~\cite{Braaten:2015tga}.
The 1-loop integrals $J_1$ and $I_1$
defined by Eqs.~\eqref{Jn} and \eqref{In} have single poles in $d-2$.
In the NLO pair propagator, these integrals appear in the combinations
$J_1(E)^2$ and $I_1(E_*)J_1(E)$, which have double and single poles.
The pole and the constant term in the two integrals are
\begin{subequations}
\bea
J_1(E) &\longrightarrow&
-\frac{1}{2 \pi} \Lambda
\left[ \frac{1}{d-2} + \frac12 \log \frac{2 \mu (E_*- E)}{\overline{\Lambda}^2} \right],
\label{J1:d->2}
\\
I_1(E_*) &\longrightarrow&
-\frac{1}{2 \pi} \Lambda
\left[ \frac{1}{d-2} + \frac12  \left( \log \frac{2 \mu E_*}{ \overline{\Lambda}^2}+ \log r -  i \pi \right) \right].
\label{I1:d->2}
\eea
\label{J1I1:d->2}%
\end{subequations}
The momentum scale in the denominator of the logarithms is
\beq
\overline{\Lambda} =\sqrt{4 \pi} e^{-\gamma_\mathrm{E}/2} \Lambda,
\label{Lambdabar}
\eeq
where $\gamma_\mathrm{E}$ is Euler's constant.
The integral $L_0$ in Eq.~\eqref{L0} has a single pole in $d-2$:
\beq
L_0(E,p) \longrightarrow
-\frac{1}{2 \pi (d-2)}  \Lambda.
\label{L0:d->2}
\eeq
The 2-loop integral $K_{110}$  defined in 
Eq.~\eqref{K110-E:d} has double and single poles in $d-2$:
\bea
K_{110}(E) &\longrightarrow&
\frac{2r}{(4\pi)^2}  \Lambda^2
\left[ \frac{1}{(d-2)^2} 
+ \frac{1}{d-2}  
\left( \log \frac{2 \mu (E_*-E)}{\overline{\Lambda}^2} 
+ \frac12 \log r \right) \right].
\label{K10:d->2}
\eea
The 2-loop integral $K_{120}$ defined in Eq.~\eqref{K120-E:d} has a single pole
in $d-2$ proportional to $1/(E-E_*)$.  However in the NLO pair propagator,
this integral appears in the combination $K_{120}(E) - r  I_1(E_*) J_2(E)$ in which the pole cancels.

\subsection{Values at the complex  threshold}
\label{sec:threshold}

The value of the one-loop integral $J_n$ defined by Eq.~\eqref{int-Jn} 
at the complex threshold $E = E_*$ is equal to 0
by the rules of dimensional regularization,
because the integral has no momentum scale:
\beq
J_n(E_*) = 0.
\label{Jn:E*}
\eeq
Note however that if $J_n(E)$ is first analytically continued to $d=3$ 
and then evaluated at $E = E_*$, it is infrared divergent for $n \ge 3/2$.
The one-loop integral $I_n$ defined by Eq.~\eqref{int-In}  evaluated at the complex threshold 
can be obtained by setting  $E = \kappa^2_*/(2\mu)$ in Eq.~\eqref{In} 
and replacing the minus sign inside the square brackets by  the complex phase $e^{-  i \pi}$:
\beq
I_n(E_*)  = 
  \frac{ \Gamma(n-d/2)}{(4 \pi)^{d/2}\Gamma(n)} 
\Lambda^{3-d} \left[ e^{-  i \pi} \, r\, \kappa^2_* \right]^{d/2-n}.
\label{In*}
\eeq

The  two-loop integrals $K_{110}$ and  $K_{120}$
defined in Eqs.~\eqref{K110,K120-E:d} can be evaluated analytically at the complex threshold: 
\begin{subequations}
\bea
K_{110}(E_*)  &=& 
\frac{2\, \Gamma(2-d)\,  r^{d/2}}{(d-2)\,(4 \pi)^d} 
\, \Lambda^{6-2d} \, \big[ e^{-  i \pi} \kappa_*^2 \big]^{d -2},
\label{K110-E*:d}
\\
K_{120}(E_*)  &=&
\frac{ 4\, \Gamma(2-d)\,  r^{d/2}}{(4-d)\,(4 \pi)^d} 
 \,  \Lambda^{6-2d} \,\big[ e^{-  i \pi}\kappa_*^2 \big]^{d-3}.
\label{K120-E*:d}
\eea
\label{K110,K120-E*:d}%
\end{subequations}
The two-loop integral $K_{111}$ defined in Eq.~\eqref{K111-E:d}
 can be evaluated analytically at the complex threshold in terms of a  hypergeometric function:
\bea
K_{111}(E_*) &=& 
- \frac{4\, \Gamma(2-d)\,  r^{d-2}}{(d-2)\,(4 \pi)^d} \,
{}_2F_1(\mbox{$\frac12$}d-1,\mbox{$\frac12$}d-1,\mbox{$\frac12$}d;1-r)
\,  \Lambda^{6-2d} \,\big[ e^{-  i \pi}\kappa_*^2 \big]^{d-3}
\label{K111-E*:d}
\eea

\subsection{Limits as $\bm{d \to 3}$}
\label{sec:poles3}

The  one-loop integrals $J_n$ and $I_n$ have finite limits as $d\to3$.
The  integrals $J_1$ and $J_2$  are
\begin{subequations}
\bea
J_1(E) &=& - \frac{\kappa(E)}{4 \pi},
\label{J1:d=3}
\\
J_2(E) &=&  \frac{1}{8 \pi\, \kappa(E)} ,
\label{J2:d=3}
\eea
\label{J1,J2:d=3}%
\end{subequations}
where $\kappa(E)$ is defined in Eq.~\eqref{kappa-E}.
In Ref.~\cite{Braaten:2015tga}, there was  an overall sign error in the expression for $J_2(E)$.
The values of $I_1$ and $I_2$ at the complex threshold $E=E_*$ are
\begin{subequations}
\bea
I_1(E_*) &=& i \,  \frac{r^{1/2}\, \kappa_*}{4 \pi},
\\
I_2(E_*) &=&  i \, \frac{r^{-1/2} }{8 \pi \, \kappa_*}.
\label{I2:d=3}
\eea
\label{I1*:d=3}%
\end{subequations}
In Ref.~\cite{Braaten:2015tga}, there was also an overall sign error in the expression for  $I_2(E)$ at $d=3$.

The one-loop integrals $L_n$ depend on $p$ as well as on $E$.
The limits of $L_0$ and $L_1$ as $d \to 3$ are
\begin{subequations}
\bea
L_0(E,p) &=& i \,\frac{r^{1/2}}{4\pi} \sqrt{2 \mu E - p^2} ,
\label{L0:d=3}
\\
L_1(E,p) &=& i\, \frac{1}{8\pi \sqrt{1-r}\, p} 
\log \frac{\sqrt{r}\, \sqrt{2 \mu E - p^2} + i \, \kappa(E) + \sqrt{1-r}\, p}{\sqrt{r}\,  \sqrt{2 \mu E - p^2} + i\, \kappa(E) -\sqrt{1-r}\, p}.
\label{L1:d=3}
\eea
\end{subequations}

The 2-loop  integrals $K_{110}$, $K_{120}$, and $K_{111}$ all have  single poles in $d-3$
that arise from logarithmic UV divergences.
The integrals $K_{110}$ and $K_{120}$ can be evaluated analytically
by making  subtractions on the Feynman-parameter integrals in Eqs.~\eqref{K110,K120-E:d}
that can be evaluated analytically in $d$ dimensions, taking the limit $d \to 3$ in the remaining  integral,  
and then evaluating that convergent integral  in terms of logarithms.
One of the possible subtractions is the   2-loop integral at the complex threshold $E_*$.
The subtracted integrals are
\begin{subequations}
\bea
K_{110}(E) - K_{110}(E_*)&=&
- \frac{2r^{3/2}}{(4\pi)^3} 
\Bigg[ i\, 2 \pi\, \kappa_* \kappa(E) 
+ \left( \frac{1}{d-3} -2 + \frac12 \log r +\log\frac{\kappa_*^2}{\bar \Lambda^2}   - i \pi \right)  \kappa^2(E)
\nonumber \\
 &&  \hspace{-1cm}
+ \big[ \kappa_* + \kappa(E) \big]^2  \log\frac{\kappa_* + \kappa(E)}{\kappa_*}
+ \big[ \kappa_* - \kappa(E) \big]^2  \log\frac{\kappa_* - \kappa(E)}{\kappa_*}
\Bigg],
\label{K110:d=3}
\\
K_{120}(E) - K_{120}(E_*) &=&
\frac{4 r^{3/2}}{(4\pi)^3} 
\Bigg[ i \, \frac{\pi}{2}  \left( \frac{\kappa_* }{\kappa(E)}  +  \frac{\kappa(E)}{\kappa_*} \right)
\nonumber \\
 &&  \hspace{-1cm}
+  \frac {[\kappa_* + \kappa(E)]^2}{2\kappa_* \kappa(E)}  \log\frac{\kappa_* + \kappa(E)}{\kappa_*}
-  \frac {[\kappa_* - \kappa(E)]^2}{2\kappa_* \kappa(E)}  \log\frac{\kappa_* - \kappa(E)}{\kappa_*}
-1 \Bigg] .
\label{K120:d=3}
\eea
\label{K110,K120:d=3}%
\end{subequations}
In Ref.~\cite{Braaten:2015tga}, the subtraction of $K_{111}$ at the complex threshold
was reduced in the limit $d \to 3$ to  a convergent integral over a single Feynman parameter.
The subtracted integral is
\bea
K_{111}(E) - K_{111}(E_*) =
-\frac{r^{3/2}}{(4\pi)^3}
 \int_0^1 {\rm d}w\, \frac{w}{\sqrt{1-w}\, (1-w+rw^2/4)}
 \log\frac{E - w E_*}{(1-w)E_*}.
\label{K111:d=3}
\eea

\subsection{Threshold expansions}
\label{sec:threshexp}

The loop integrals $K_{lmn}$ defined in Eq.~\eqref{int-Klmn} 
have threshold expansions in powers of $E-E_*$.
In the limit $d\to 3$, these expansions are in half-integer powers of $E-E_*$
or, equivalently, in integer powers of the function $\kappa(E)$ defined in Eq.~\eqref{kappa-E}.
The threshold expansions for $K_{110}$ and $K_{120}$ can be obtained using Eqs.~\eqref {K110,K120-E*:d}
and \eqref{K110,K120:d=3}.
The threshold expansions through second order in $E-E_*$ are
\begin{subequations}
\bea
K_{110}(E) &=&
- \frac{2r^{3/2}}{(4\pi)^3}\, \kappa_*^2
\Bigg[ \left( \frac{1}{d-3} - 2 +\frac12 \log r
+\log \frac{\kappa_*^2}{\overline{\Lambda}^2}  -  i \pi \right) + i\, 2 \pi\, \frac{\kappa(E)}{\kappa_*}
\nonumber \\
 &&  \hspace{2cm}
+ \left( \frac{1}{d-3} +1 + \frac12 \log r +\log\frac{\kappa_*^2}{\bar \Lambda^2}   - i \pi \right)  \frac{\kappa^2(E)}{\kappa_*^2}
\nonumber \\
 &&  \hspace{2cm}
- \frac{1}{6}\,\frac{\kappa^4(E)}{\kappa_*^4} + {\cal O}\big( \kappa^6(E) \big)
\Bigg],
\label{K110:thresh}
\\
K_{120}(E) &=&
\frac{4 r^{3/2}}{(4\pi)^3} 
\Bigg[ \left( \frac{1}{d-3} +\frac12 \log r
+\log \frac{\kappa_*^2}{\overline{\Lambda}^2}  -  i \pi \right) + i \, \frac{\pi}{2}  \left( \frac{\kappa_* }{\kappa(E)}  +  \frac{\kappa(E)}{\kappa_*} \right)
\nonumber \\
 &&  \hspace{2cm}
+ \frac{1}{3}\,\frac{\kappa^2(E)}{\kappa_*^2} + \frac{1}{30}\,\frac{\kappa^4(E)}{\kappa_*^4} + {\cal O}\big( \kappa^6(E) \big) \Bigg].
\label{K120:thresh}
\eea
\end{subequations}
In Ref.~\cite{Braaten:2015tga}, the $\kappa^2(E)$ terms  have the wrong sign.
A method to obtain the threshold expansion for $K_{111}$ 
was presented in Ref.~\cite{Braaten:2015tga}.
The coefficients can be expressed analytically in terms of hypergeometric functions.
The threshold expansion through second order in $E-E_*$ is
\bea
K_{111}(E) &=& -\frac{4}{(4 \pi)^3}\bigg[ 
\left( \frac{1}{d-3} - 2 + \log r 
+ \log \frac{\kappa_*^2}{\overline{\Lambda}^2}  -  i \pi \right) \frac{r}{ \sqrt{1-r}} \arccos \big(\sqrt{r}\,\big)
\nonumber\\
&& \hspace{4cm} 
+r\, \frac{\rm{d}\ }{{\rm d}d}\,{}_2F_1(\mbox{$\frac12$}d-1,\mbox{$\frac12$}d-1,\mbox{$\frac12$}d;1-r)
\Big |_{d=3} 
\nonumber\\
&& \hspace{2cm}
+  \left( \frac{2-r}{\sqrt{1-r}} \arccos \big(\sqrt{r}\,\big) + 2 \sqrt{r} \right)
\frac{\kappa^2(E)}{\kappa_*^2}
\nonumber\\
&& \hspace{2cm}
-\frac{4}{r}  
\bigg(  {}_2F_1\big(\mbox{$-\frac12$},\mbox{$-\frac12$},\mbox{$\frac32$},1-r\big) 
+ \frac{2-4r+r^2}{8\sqrt{1-r}} \arccos \big(\sqrt{r}\,\big) 
\nonumber \\
&& \hspace{4cm}
+  \frac{3-4r}{12 }\sqrt{r} \bigg)
\frac{\kappa^4(E)}{\kappa_*^4}
+ {\cal O}\big( \kappa^6(E) \big) \bigg]. 
\label{K111:thresh}
\eea
In Ref.~\cite{Braaten:2015tga}, the $\kappa^2(E)$ term has the wrong sign.

\section{Diagrams for NLO Transition Amplitudes}
\label{sec:NLOAmp}

In this Appendix, we give the results for  the NLO Feynman diagrams for 
the amputated connected Green functions for 
the transitions $D^{*0} \bar D^0,D^0 \bar D^{*0} \to D^{*0} \bar D^0,D^0 \bar D^{*0}$.
Most of the diagrams were evaluated in Appendix~B of Ref.~\cite{Braaten:2015tga}.
We correct errors in some of the pion-exchange diagrams in Ref.~\cite{Braaten:2015tga}.
We calculate  the NLO diagrams
using the new Feynman rules in Section~\ref{sec:NewVertices}. 
We verify that the sum of the diagrams is the same as that calculated using 
the Feynman rules in Ref.~\cite{Braaten:2015tga}.

\subsection{Leading order diagram}

The amputated connected Green function for $D^{*0} \bar D^0 \to D^{*0} \bar D^0$ is a tensor $i\, \mathcal{T}^{ij}$
whose vector indices are  those of the incoming and outgoing $D^{*0}$ lines.
We set the incoming and outgoing spin-0 charm mesons on their energy shells.
The transition tensor $\mathcal{T}^{ij}(E, \bm{p},\bm{p}^\prime)$ is then
a function of the total energy $E$ of the charm mesons in their CM frame
and the relative momenta $\bm{p}$ and $\bm{p}^\prime$ of the incoming and outgoing spin-0 charm mesons.
The S-wave contribution to the  transition tensor can be obtained by averaging 
over the directions of $\bm{p}$ and $\bm{p}^\prime$.

The diagram for the transition tensor $\mathcal{T}^{ij}$ for $D^{*0} \bar D^0 \to D^{*0} \bar D^0$ at LO
is the right side of Fig.~\ref{fig:vertex-pairprop}.
This transition tensor is diagonal in the vector indices, and it is a function of $E$ only.
The LO  transition tensor for $D^{*0} \bar D^0 \to D^{*0} \bar D^0$ is
\beq	
\mathcal{T}^{ij}_\mathrm{LO}(E) = \left(\frac{1}{\sqrt2}\right)^2 \frac{2 \pi}{\mu} \mathcal{A}(E) \, \delta^{ij}.
\label{Tij-LO}
\eeq

\subsection{Pion-exchange diagrams}

\begin{figure}[tb]
\centerline{\includegraphics*[width=16cm,angle=0,clip=true]{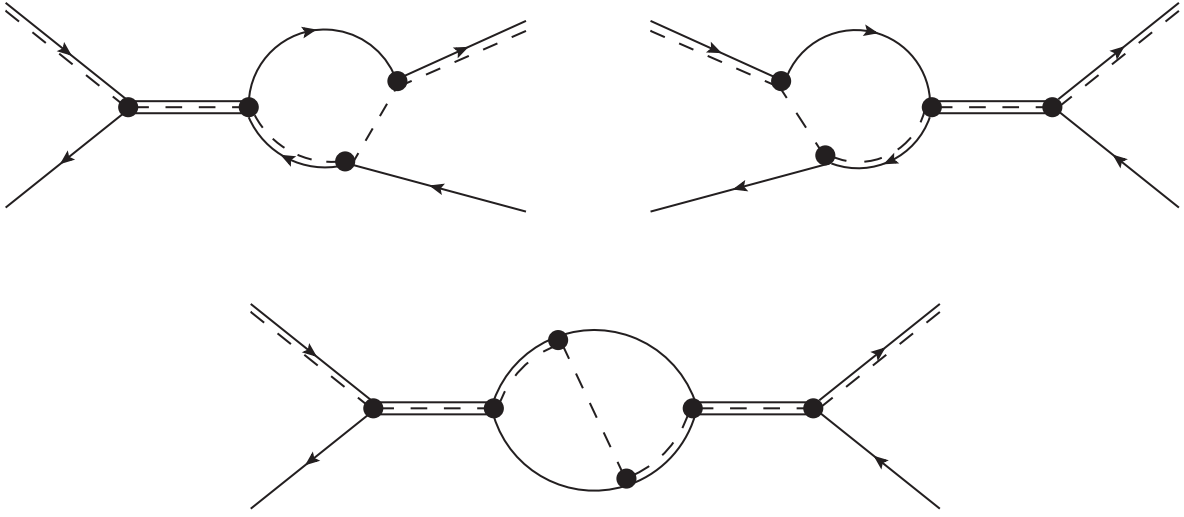}}
\vspace*{0.0cm}
\caption{
The pion-exchange diagrams for $D^{*0} \bar D^0 \to D^{*0} \bar D^0$
consist of the 1-loop diagrams A1 and A2 and the 2-loop diagram A3. 
The absence of arrows on the charm-mesons lines in the loop for the diagram A3 implies 
a sum over the two possible directions of the arrows.
}
\label{fig:NLOpiex}
\end{figure}

There are three NLO pion-exchange diagrams for $D^{*0} \bar D^0 \to D^{*0} \bar D^0$.
They are shown in Fig.~\ref{fig:NLOpiex} and are labelled 
A1, A2, and A3.

The amplitudes for the 1-loop diagrams A1 and A2 in Fig.~\ref{fig:NLOpiex} 
are tensors in the indices $i$ and $j$ that depend on the relative momenta $\bm{p}'$ and $\bm{p}$, respectively.
The amplitude for the diagram A2 is
\bea
\mathcal{T}_\mathrm{A2}^{ij}(E,\bm{p})  &=& 
\frac{r \pi g_\pi^2/\mu}{(d-1)\sqrt{1-r}\, p^2} \mathcal{A}(E) \Big\{ \big[ (\Delta(E,p) -2 r p^2 ) L_0(E,p)
\nonumber
\\
&& \hspace{5cm}
- (\Delta(E,p)^2 -8 r \mu E_* p^2) L_1(E,p)
\nonumber
\\
&& \hspace{5cm}
-  (\Delta(E,p) - 2 p^2 ) J_1(E) \big]
( \delta^{ij} - p^i p^j/p^2 )
\nonumber
\\
&& \hspace{1cm}
-  (d-1)  \big[ \Delta(E,p)  L_0(E,p)
- \Delta(E,p) (\Delta(E,p) -2 r p^2 ) L_1(E,p)
\nonumber
\\
&& \hspace{5cm}
- (\Delta(E,p) +2 (1-r) p^2 ) J_1(E) \big] p^i p^j/p^2  \Big\} ,
\label{A3}
\eea
where $r = \mu_\pi/\mu$ and $\Delta$ is a linear function of $E$ and $p^2$:
\beq	
\Delta(E,p) =2 \mu E_*  - 2(1-r)\mu E + p^2.
\label{DeltaE}
\eeq
The S-wave contribution of the diagram can be obtained by averaging over the directions of $\bm{p}$
in $d$ spatial dimensions:
\bea
\Big\langle \mathcal{T}^{ij}_\mathrm{A2}(E,\bm{p}) \Big\rangle_{\hat{\bm{p}}} &=& 
\frac{2\pi  r g_\pi^2/\mu}{d \sqrt{1-r}}
\big[- r\, L_0(E,p) - r\, \big( \Delta(E,p) - 4  \mu E_*\big)L_1(E,p) 
\nonumber\\
&& \hspace{2cm}+  (2-r) J_1(E) \big] 
 \mathcal{A}(E) \, \delta^{ij}.
\label{AS,3}
\eea
The S-wave contribution $\langle \mathcal{T}^{ij}_\mathrm{A1}(E,\bm{p}^\prime) \rangle_{\hat{\bm{p}}^\prime}$
to the diagram A1 can be obtained  by replacing $p$ in Eq.~\eqref{AS,3} by $p^\prime$.
In Ref.~\cite{Braaten:2015tga},
it was stated incorrectly that the $p^ip^j/p^2$ terms in Eq.~\eqref{A3} cancel in the limit $\bm{p} \to 0$.
They do however cancel in the limit $p\to 0$ after averaging over the directions of $\bm{p}$.

The 2-loop diagram A3  in Fig.~\ref{fig:NLOpiex} is the sum of a diagram 
with a $D^{*0}\bar D^0$ loop and a diagram with a $D^0\bar D^{*0}$ loop.
The amplitude for this diagram is 
\bea	
\mathcal{T}^{ij}_\mathrm{A3}(E)  =
- \frac{8 \pi^2 r g_\pi^2/\mu}{d \sqrt{1-r}} 
\big[ 2 K_{110}(E) -2 \mu(2  E_* - r  E)K_{111}(E)  
- (2-r) J_1(E)^2 \big] \mathcal{A}^2(E) \, \delta^{ij}.~~~
\label{A4}
\eea
%

\begin{figure}[tb]
\centerline{\includegraphics*[width=6cm,angle=0,clip=true]{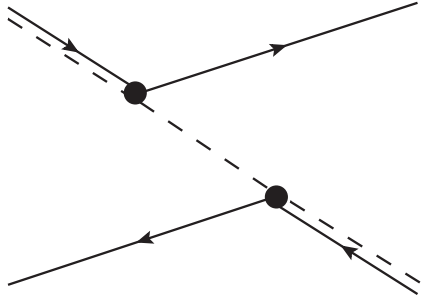}}
\vspace*{0.0cm}
\caption{
The pion-exchange tree diagram for $D^{*0} \bar{D}^0 \to D^0 \bar D^{*0}$, which is labelled $\pi$.
There are also three  loop diagrams that can be obtained from those 
in Fig.~\ref{fig:NLOpiex} by changing the directions of some of the arrows.
}
\label{fig:NLOpiextree}
\end{figure}

For each of the pion-exchange diagrams for $D^{*0} \bar D^0 \to D^{*0} \bar D^0$ in Fig.~\ref{fig:NLOpiex},
there is an analogous diagram for $D^{*0} \bar D^0 \to D^0 \bar D^{*0}$,
$D^0 \bar D^{*0} \to D^{*0} \bar D^0$, and $D^0 \bar D^{*0} \to D^0 \bar D^{*0}$.
For $D^{*0} \bar D^0 \to D^0 \bar D^{*0}$ and $D^0 \bar D^{*0} \to D^{*0} \bar D^0$,
there is also a tree diagram.
The tree diagram for $D^{*0} \bar{D}^0 \to D^0 \bar D^{*0}$
 is shown in Fig.~\ref{fig:NLOpiextree}, and it is labelled $\pi$.
The transition tensor  for  this diagram depends on the relative momenta $\bm{p}$ and $\bm{p}'$:
\bea
\mathcal{T}_{\pi}^{ij}(E,\bm{p},\bm{p}')  = 
- \frac{g_\pi^2/\mu^2}{E - (p^2 +{p'}^2)/(2M) - (\bm{p} + \bm{p}')^2/(2m) + i \epsilon}
\nonumber
\\
\times \left( \frac{M}{M+m} \bm{p} + \bm{p}' \right)^i
\left( \frac{M}{M+m} \bm{p}' + \bm{p} \right)^j.
\label{A1}
\eea
This diagram vanishes in the zero-momentum limit $\bm{p},\bm{p}' \to 0$ with $E$ fixed.
Since this diagrams does not have any UV divergences,
the S-wave contribution of the diagram can be obtained by averaging over the directions 
of $\bm{p}$ and $\bm{p}^\prime$  in 3 spatial dimensions:
\bea
\Big\langle \mathcal{T}^{ij}_{\pi}(E,\bm{p},\bm{p}') \Big\rangle_{\hat {\bm{p}},\hat{\bm{p}}^\prime} &=& 
-  \frac{r g_\pi^2/\mu}{3  \sqrt{1-r}} \Bigg( \frac{ r [ 2 (2-r) \mu E - p^2- p^{\prime 2}]}{4 \sqrt{1-r} \, p \,p^\prime}
\nonumber\\
&& \hspace{1cm}
\times \log \frac{2 r \mu E - p^2- p^{\prime 2} +2 \sqrt{1-r}\,  p\,p^\prime}
{2 r \mu E - p^2 - p^{\prime 2} - 2 \sqrt{1-r}\,  p \,p^\prime}
- (2-r) \Bigg) \delta^{ij}.
\label{Tpi-S}
\eea

\subsection{$\bm{D^*}$ propagator correction diagrams}

\begin{figure}[tb]
\centerline{\includegraphics*[height=6cm,angle=0,clip=true]{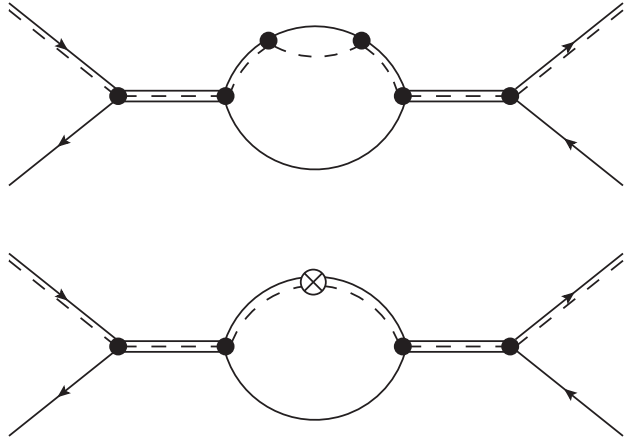}}
\vspace*{0.0cm}
\caption{
The $D^*$ propagator correction diagrams for $D^{*0} \bar D^0 \to D^{*0} \bar D^0$
consist of the 2-loop diagram B1 with a $D^*$ self-energy subdiagram 
and the 1-loop diagram B2 with a $D^*$ self-energy counterterm.
}
\label{fig:NLOD*se}
\end{figure}

The two NLO diagrams involving  $D^*$ propagator corrections
are shown in Fig.~\ref{fig:NLOD*se} and are labelled B1 and B2.
The diagram B1, which  has a $D^*$ self-energy subdiagram 
inserted into the $D^*$ propagator, can be reduced to
\beq
\mathcal{T}^{ij}_\mathrm{B1}(E)  =
- \frac{16 \pi^2 r g_\pi^2/\mu}{d} \big[ K_{110}(E) - 2 \mu E_* K_{120}(E) \big]
\mathcal{A}^2(E) \, \delta^{ij}.
\label{B1}
\eeq
The diagram B2, which has a $D^*$ propagator counterterm 
inserted into the $D^*$ propagator, can be reduced to
\beq
\mathcal{T}^{ij}_\mathrm{B2}(E)  =
-\frac{8 \pi^2 r^2 g_\pi^2/\mu}{d} I_1(E_*) \big[ 4 \mu E_* J_2(E) - d J_1(E)  \big]
\mathcal{A}^2(E) \, \delta^{ij}.
\label{B2}
\eeq

\subsection{$\bm{\nabla^2}$ vertex diagrams}

\begin{figure}[tb]
\centerline{\includegraphics*[width=12cm,angle=0,clip=true]{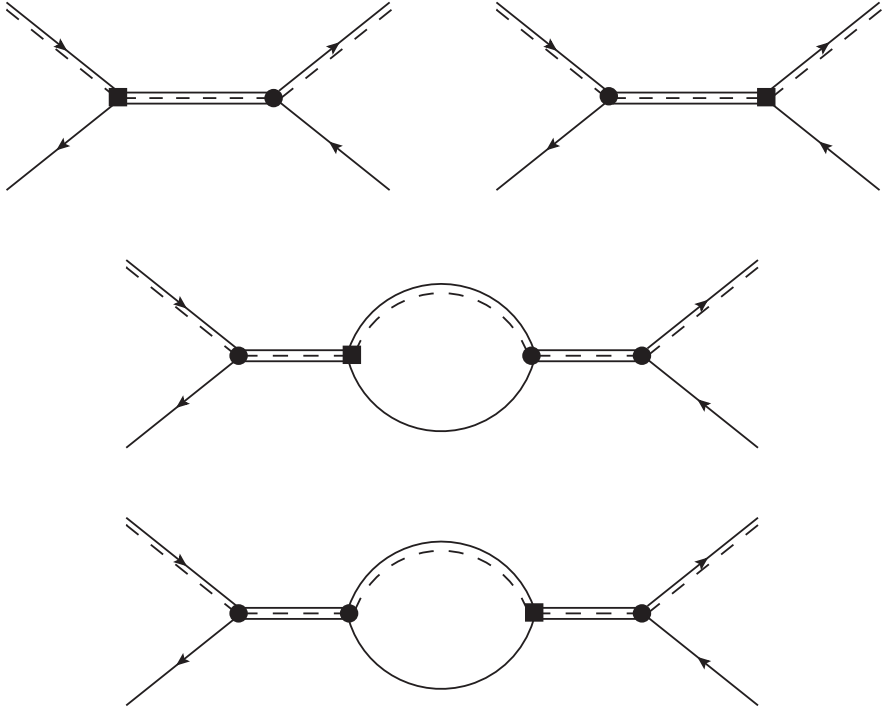}}
\vspace*{0.0cm}
\caption{
The  $\nabla^2$ vertex diagrams for $D^{*0} \bar D^0 \to D^{*0} \bar D^0$
 consist of two tree diagrams C1 and C2
and two 1-loop diagrams  C3 and C4.
}
\label{fig:NLOnabla2:3}
\end{figure}

The four NLO diagrams with a $\nabla^2$ vertex
are shown in Fig.~\ref{fig:NLOnabla2:3} and are labelled C1, C2, C3, and C4.
The diagram C1 in which the incoming 
charm mesons interact through a $\nabla^2$ vertex is
\beq	
\mathcal{T}_\mathrm{C1}^{ij}(E,p)  =  \frac{\pi C_2}{2 \mu C_0}\mathcal{A}(E)  \, p^2\, \delta^{ij}.
\label{C1-new}
\eeq
The diagram C2 in which the outgoing 
charm mesons interact through a $\nabla^2$ vertex is obtained by replacing $p^2$ by ${p'}^2$.
The diagram C3 is
\beq	
\mathcal{T}_\mathrm{C3}^{ij}(E)  = 
\frac{4 \pi^2 C_2}{C_0} (E-E_*) J_1(E)\, \mathcal{A}^2(E)  \, \delta^{ij}.
\label{C3-new}
\eeq
The expression for the diagram C4 is the same.
The sum of the four diagrams  is
\beq
\mathcal{T}_{\mathrm{C}}^{ij}(E,p,p')  = 
\frac{\pi C_2}{2 \mu C_0} \Big[ (p^2 + {p'}^2) \mathcal{A}(E) 
+ 16 \pi \mu (E - E_*) J_1(E) \mathcal{A}^2(E)  \Big] \delta^{ij}.
\label{C1234-new}
\eeq

With the Feynman rules in Ref.~\cite{Braaten:2015tga}, the $\nabla^2$ vertices are $2 \to 2$ vertices.
There are four $\nabla^2$ vertex diagrams:  the $\nabla^2$ vertex,
two 1-loop diagrams with a single factor of $\mathcal{A}(E)$,
and a 2-loop diagram with two  factors of $\mathcal{A}(E)$.
The  identity in Eq.~\eqref{identity} can be used to show
 the sum of the four diagrams is the same as that in Eq.~\eqref{C1234-new}.

\subsection{Pair-propagator counterterm diagram}

\begin{figure}[tb]
\centerline{\includegraphics*[width=8cm,angle=0,clip=true]{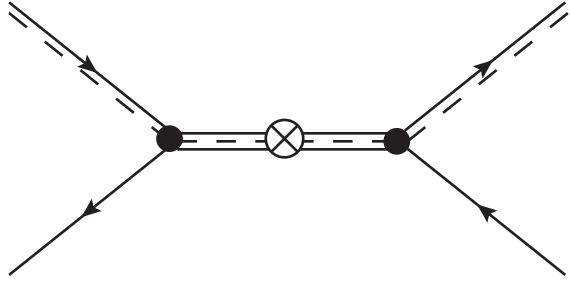}}
\vspace*{0.0cm}
\caption{
The pair propagator  counterterm diagram for $D^{*0} \bar D^0 \to D^{*0} \bar D^0$,
which is labelled D.
}
\label{fig:NLOcount:3}
\end{figure}

The  NLO diagram with a pair-propagator counterterm is shown in Fig.~\ref{fig:NLOcount:3}.
The expression for the  diagram is
\beq	
\mathcal{T}^{ij}_{\mathrm{D}}(E)  =
-\frac{2 \pi^2 }{\mu^2 C_0^2}  \, \big[ \delta C_0 +D_0 \, (E-E_*)\big]\,
\mathcal{A}^2(E) \delta^{ij}.
\label{D-new}
\eeq

With the Feynman rules in  Ref.~\cite{Braaten:2015tga}, there is a $2 \to 2$ vertex counterterm.
There are four NLO diagrams with a counterterm vertex:  the vertex counterterm,
two 1-loop diagrams with a single factor of $\mathcal{A}(E)$,
and a 2-loop diagram with two  factors of $\mathcal{A}(E)$.
The  identity in Eq.~\eqref{identity} can be used to show
 the sum of the four diagrams is equal to that in Eq.~\eqref{D-new}.

\end{appendix}

\end{document}